\documentclass[10pt,letterpaper,onecolumn]{article}
\pdfoutput=1
\usepackage[utf8]{inputenc}
\usepackage[english]{babel}
\usepackage{amsmath}
\usepackage{amsfonts}
\usepackage{amssymb}
\usepackage{graphicx}
\usepackage{multirow}
\usepackage{url}
\usepackage[numbers]{natbib}

\title{Automatic Temperature Setpoint Tuning of a Thermoforming Machine using Fuzzy Terminal Iterative Learning Control}
\author{Mathieu Beauchemin-Turcotte$^*$, Guy Gauthier\footnote{Affiliated with Department of automated engineering, École de technologie supérieure, Montréal, Canada.}\footnote{Corresponding author: guy.gauthier@etsmtl.ca} \\ and Robert Sabourin$^*$}
\begin{document}
\maketitle

\section*{Abstract}
This paper presents a new way to design a Fuzzy Terminal Iterative Learning Control (TILC) to control the heater temperature setpoints of a thermoforming machine. This fuzzy TILC is based on the inverse of a fuzzy model of this machine, and is built from experimental (or simulation) data with kriging interpolation. The Fuzzy Inference System usually used for a fuzzy model is the zero order Takagi Sugeno Kwan system (constant consequents). In this paper, the 1st order Takagi Sugeno Kwan system is used, with the fuzzy model rules expressed using matrices. This makes the inversion of the fuzzy model much easier than the inversion of the fuzzy model based on the TSK of order 0. Based on simulation results, the proposed fuzzy TILC seems able to give a very good initial guess as to the heater temperature setpoints, making it possible to have almost no wastage of plastic sheets. Simulation results show the effectiveness of the fuzzy TILC compared to a crisp TILC, even though the fuzzy controller is based on a fuzzy model built from noisy data.

\section*{Keywords}
Fuzzy internal model control, fuzzy modeling, kriging interpolation, terminal iterative learning control. 

\section{Introduction}
The thermoforming process plays an important role in the plastics industry. In North America in 2003, in excess of \$10 billion worth of thermoformed parts were produced \citep{Throne2002}. Examples of the parts fabricated using this process include car bumpers, baths, and packaging products, among numerous others. 

The plastic sheets used in the thermoforming process undergo three phases \cite{Throne1996,Florian1987,Ajersch2004}. The reheat phase, where the plastic sheet is heated to a sufficiently high temperature to render it malleable; the molding phase; and the cooling phase. The malleable plastic sheet is draped over a mold to give it the desired shape, and then cooled in the mold until it becomes rigid again. Finally, in the trimming phase, excess plastic is removed to obtain a usable part. 

In this paper, we address only issues arising in the reheat phase of the thermoforming process. This phase is critical, since the molding will not be accurate if the plastic sheet is not heated correctly \cite{Florian1987,Chy2010}, resulting in rejection of the inaccurately molded part by the quality control department.

To heat the plastic sheet correctly, the heater temperature setpoints have to be appropriately adjusted. This adjustment is performed by the operator of the machine using a trial-and-error approach. This approach results in plastic sheets being wasted, and occurs mostly at the start of a new production lot. It also occurs when the temperature in the plant changes significantly after the adjustment is made. In North America, plastic sheet wastage during the manufacture of a component can represent up to 10 to 15\% of each production lot. For complicated parts, this percentage can rise to 20\% \cite{Florian1987}.

An experienced operator can successfully reduce these losses. However, plastics processing companies are struggling to find qualified and experienced people to operate their thermoforming machines. Their fear now is that a large number of these employees will reach retirement age over the next few years.

Very little work has been published regarding the control of sheet surface temperature \cite{Girard2005}. Such a control system would improve the quality of the parts produced, reduce scrap, and allow for temperature zoning. Plastic sheet temperature is controlled by adjusting the heater temperature of the thermoforming machine, and this control must be such that the plastic sheet temperature converges to a desired temperature profile. This issue belongs to a class of problems known as ”inverse heating problems” (IHP) \cite{Duarte2002}.

Existing TILC algorithms are highly dependent on initial guesswork with regard to temperature setpoints. When the initial guess is near the temperature setpoint required to heat a plastic sheet correctly, all is well. But, when it is not, a number of cycles are required to converge to the right value, causing wastage of plastic sheets. By increasing the controller gain, the convergence speed can be improved, but at the cost of less robustness. Furthermore, with those high gains, the system can be subject to overshoot. For a thermoforming oven, this is a serious problem, since there is a risk that a very malleable, overheated plastic sheet will fall on the bottom heaters, forcing the process to be stopped in order to clean them.

A suitable approach that makes this initial guess reliably would be a major improvement in the control of this kind of process, where money needs to be saved by reducing wastage. It would result in the right heater temperature setpoints being determined at the start of the process, and reduce the controller gain perturbations, improving robustness.

The selected control approach is a TILC based on an inverse fuzzy model. This nonlinear control approach permits a relatively good initial guess of heater temperature setpoints from a desired surface temperature profile of the plastic sheet. The design approach presented here provides the opportunity to use data from a reliable nonlinear model, or from the process itself. This fuzzy approach has never been used for a TILC. The main contribution of this paper is the use of the 1st order Takagi Sugeno Kwan Fuzzy Inference System, along with its rule consequent, expressed in matrix form and built from kriging interpolation.

In this paper, we make a number of assumptions: 1) That the ambient temperature is constant (this assumption is relaxed in the simulations results); 2) That the initial temperature of the plastic sheet is constant; 3) That the plastic sheet is homogeneous and not subject to sag; 4) That the PID loops, in which the setpoints interact with the heater \cite{Ajersch2004}, have reached their steady state when the plastic sheet is heated; 5) That the number of inputs is equal to the number of outputs; and 6) That the minimal and maximal output values are located at opposite corners of the input space. This last assumption fits well with the way the thermoforming oven functions, since the plastic sheet surface temperature is minimal when all the heaters are at minimal temperature, and maximal when all the heaters are at maximal temperature.

The paper is structured as follows. Related works are covered in section \ref{se:Related}. A system overview is provided in section \ref{se:Overview}. Section \ref{se:Modeling} explains how the fuzzy model of a nonlinear process is obtained. Section \ref{se:Inversion} shows the steps to follow to invert the fuzzy model, giving the inverse fuzzy model component of the TILC (Figure \ref{fig:Proposed_TILC}). Section \ref{se:Filter} concerns the fuzzy filter used in the fuzzy TILC that adjusts the heater temperature setpoints of the thermoforming oven. Section \ref{se:Simulation} presents the case of a fuzzy TILC applied to a six heater, six sensor thermoforming oven model, and simulation results demonstrate how the fuzzy TILC behaves. Finally, we provide our conclusions in section \ref{se:Conclusion}.

\section{Related works}\label{se:Related}

\subsection{Introducing Terminal Iterative Learning Control}
Analysis of this temperature control problem gave us the idea of using a cycle-to-cycle approach to automatically tune the heater temperature setpoints, and led to considering a Terminal Iterative Learning Control (TILC). This approach was first introduced in \cite{Chen1997b} and then in a Ph.D. thesis  \cite{Chen1997}, and is derived from Iterative Learning Control (ILC) \cite{Chen1997,Moore1999}. It has been used for the cycle-to-cycle control of a Rapid Thermal Process and Chemical Vapor Deposition (RPTCVD)  \cite{Chen1997b,Chen1997,Chen1999}.  The main difference between ILC and TILC is that ILC uses measurements sampled along the entire cycle, while TILC only uses measurements sampled at the end of the cycle.

The TILC is based on certain assumptions  \cite{Chen1997,Moore1999,Gauthier2008}:
\begin{itemize}
\item That the cycle length $T \in \mathbb{R}$ has a fixed duration;
\item That the input vector $\mathbf{u}[k] \in \mathbb{R}^m$ is maintained constant during the entire cycle $k$ ($k \in \mathbb{N}$);
\item That the process parameters can vary in the time domain ($t$) but are invariant in the cycle domain ($k$);
\item That the process start always with the same initial state vector $\mathbf{x}_0 \in \mathbb{R}^n$, so $\mathbf{x}_0[k] = \mathbf{x}_0 \:\; \forall k$;
\item That the desired output vector $\mathbf{y}_d \in \mathbb{R}^p$ must be feasible, so an input vector $\mathbf{u}[k]  \in \mathbb{R}^m$ must exist that made the terminal output $\mathbf{y}_T[k]  \in \mathbb{R}^p$ equal to $\mathbf{y}_d$; and, finally,
\item That this input vector $\mathbf{u}[k]$ must be unique.
\end{itemize}

These assumptions fit relatively well with the thermoforming reheat phase. To use the TILC approach on a thermoforming machine, temperature sensors are installed to measure the surface temperature of the plastic sheet at the end of the heating cycle  \cite{Ajersch2004,Girard2005,Gauthier2008}. TILC iteratively adjusts the heater temperature setpoints, so that the sheet surface temperature converges to the desired level at the end of the heating cycle  \cite{Gauthier2008}.

\begin{figure*}[t]
	\centering	\includegraphics[width=\textwidth]{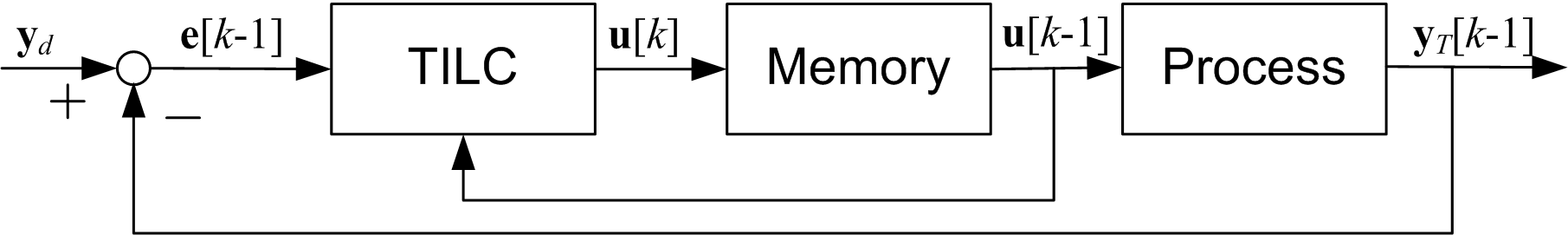}
	\caption{Block diagram of a TILC controlled system}
	\label{fig:TILC_block}
\end{figure*}

The block diagram of a TILC algorithm is shown in Figure \ref{fig:TILC_block}. The input vector $\mathbf{u}[k]$, applied to the process during the entire cycle $k$, is a function of the terminal output vector  $\mathbf{y}_T[k-1]$ measured at the end of the previous cycle $(k-1)$, the corresponding input vector $\mathbf{u}[k-1]$ and the desired terminal output vector $\mathbf{y}_d$. The simplest TILC algorithm (called a 1st order TILC) is expressed as follows:  
\begin{equation}
	\mathbf{u}[k] = \mathbf{u}[k-1] + \mathbf{K}(\mathbf{y}_d-\mathbf{y}_T[k-1])
\end{equation}
where $\mathbf{K} \in \mathbb{R}^{m \times p}$ is the matrix containing the TILC gains.

\subsection{Other related control approaches}
This problem has been addressed in semiconductor industry processes like the Rapid Thermal Process and Chemical Vapor Deposition (RPTCVD), by the use of run-to-run (R2R) control. Like TILC, the R2R algorithm uses only measurements performed at the end of the cycle or later  \cite{Zhang2000,Moyne2001,Adivikolanu1998,Sachs1995}. The R2R approach is composed of various control algorithms  \cite{Moyne2001}, the one most often used being the Exponential Weighting Moving Average (EWMA) \cite{Zhang2000,Moyne2001}. R2R control is an approach derived from statistical process control.

Papers \cite{Duarte2002,Duarte2003} cover control of the temperature profile of a plastic material in roll-fed thermoforming. This control approach is based on a computational finite element algorithm, which is mainly based on the geometric relationship between the heating elements and the plastic material. These relationships are used to compute the heater temperature setpoints. In \cite{Duarte2002}, this inverse problem is considered to be an ill-posed problem, since two sets of heater temperature setpoints can give relatively similar surface temperature profiles. In \cite{Duarte2003}, the sensitivity of the algorithm proposed in \cite{Duarte2002} to perturbations is studied.

In a recent paper \cite{Chy2010b}, the plastic sheet temperature is controlled with a two-dimensional fast Fourier transform (FFT) of the sheet temperature profile. This approach is also explained in a book chapter \cite{Chy2011}. These authors have also proposed an alternative approach, based on the conjugate gradient method \cite{Chy2010}.

\section{System overview}\label{se:Overview}

\begin{figure*}[t]
	\centering	\includegraphics[width=\textwidth]{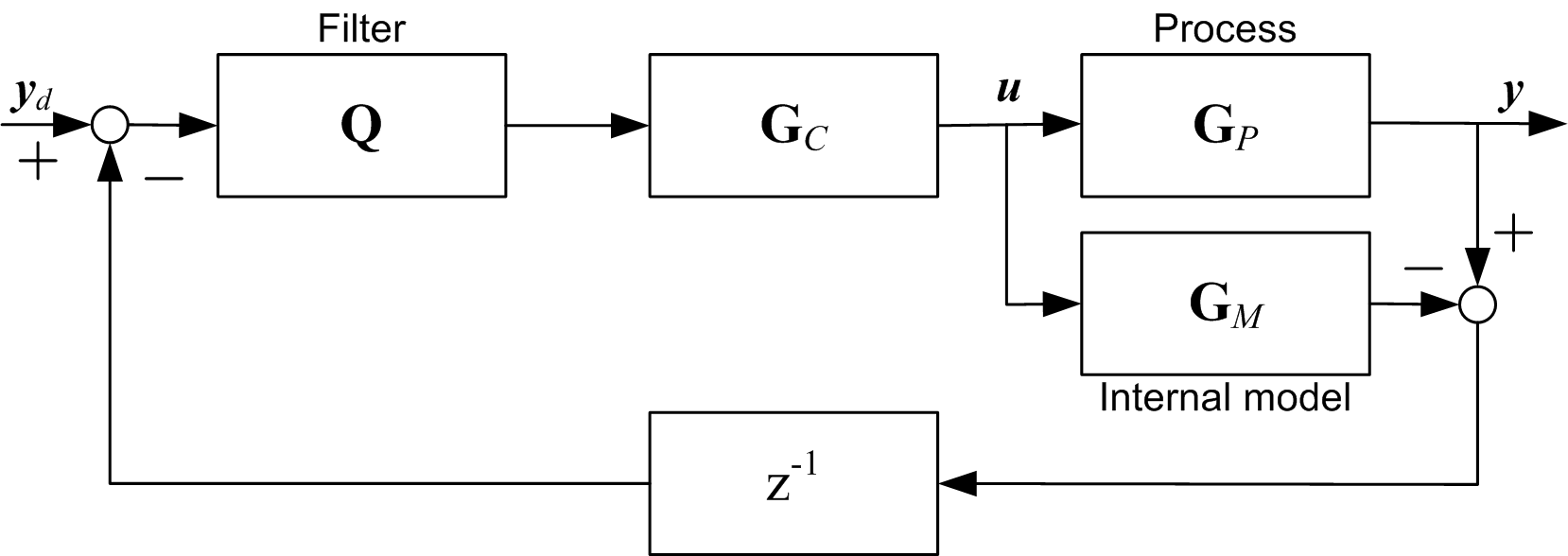}
	\caption{Internal model control structure (non-fuzzy)}
	\label{fig:IMC_block01}
\end{figure*}

The fuzzy TILC (f-TILC) proposed in this paper is based on a TILC using a fuzzy nonlinear internal model control (FNIMC). The non fuzzy TILC with Internal Model Control (IMC) proposed in the chapter 3 of \cite{Gauthier2008} has the structure shown in Figure \ref{fig:IMC_block01} (where $\mathbf{G}_P$ is the process, $\mathbf{G}_M$ is the process model, $\mathbf{G}_C$ is the controller and $\mathbf{Q}$ is the filter). 

\begin{figure*}[htbp]
	\centering	\includegraphics[width=\textwidth]{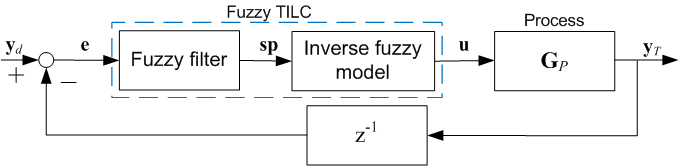}
	\caption{Proposed fuzzy TILC with FNIMC structure}
	\label{fig:Proposed_TILC}
\end{figure*}

The block diagram of the proposed fuzzy TILC with IMC is shown in Figure \ref{fig:Proposed_TILC}. This block diagram looks similar to the FNIMC block diagram shown in Figure 3 of  \cite{Boukezzoula2003}, except for the $z^{-1}$ block in the feedback. 

In the block diagram of the proposed fuzzy TILC with IMC, the filter is also a fuzzy function that permits the TILC to exhibit a behavior, such that the convergence of the output vector $\mathbf{y}_T[k]$ to $\mathbf{y}_d$ is relatively monotonic. It is used to evaluate the corrected setpoint for the next cycle $\mathbf{sp}[k]$, which in turn is used to evaluate the input $\mathbf{u}[k]$. The initial value $\mathbf{sp}[0]$ is set equal to the desired value $\mathbf{y}_d$.

The fuzzy internal model control approach has been covered by a few researchers who have considered the Takagi-Sugeno fuzzy model with constant consequents \cite{Boukezzoula2003,Boukezzoula2001,Salman2007,Galichet2004,Fu2008}.

In \cite{Boukezzoula2001}, the modelized nonlinear system with fuzzy logic is a multiple input single output (MISO) system that is converted into a single input single output (SISO) system before being inverted. This inverted fuzzy model is used in a Fuzzy Nonlinear Internal Model Control (FNIMC).

This approach was extended in \cite{Boukezzoula2003} to a multiple input multiple output (MIMO) process. These authors consider a class of MIMO processes that can be decomposed into MISO subsystems, each modeled using fuzzy logic as in \cite{Boukezzoula2001}. The Fuzzy Inference System (FIS) used for the fuzzy model is the Takagi-Sugeno-Kwan (TSK) FIS of order 0.

In \cite{Galichet2004}, the inversion of a fuzzy model is presented. Some of the mathematical notations used in this paper were inspired to some extent by \cite{Galichet2004}. In \cite{Fu2008}, a nonlinear process is decomposed using a fuzzy decoupling controller for a nuclear power plant.

Since the thermoforming process is a nonlinear MIMO system that can be decomposed into MISO subsystems, the approach presented in \cite{Boukezzoula2003} is applicable. In this paper, the FIS used is the TSK of order 1, and the output of each rule is a linear polynomial. At the defuzzification stage, the combination of those linear rules with membership functions generates nonlinear outputs. The approach used for the SISO fuzzy model inversion is extended to MIMO fuzzy model inversion by expressing the rules in matrix form. The rules of the inverse of the model are calculated using a matrix calculation. A 1st order TSK FIS for the fuzzy model provides an easier way to invert the fuzzy model than the TSK FIS of order 0, especially for large MIMO systems. 
 
The design of the proposed fuzzy TILC is based on the inverse of the fuzzy model of the process (Figure \ref{fig:Proposed_TILC}). To be able to create this design, we first need to obtain the fuzzy model of the nonlinear process, which we do from experiments performed on the process, by applying inputs and measuring the corresponding outputs. The fuzzy model of the process is based on the Takagi Sugeno Kwan (TSK) fuzzy inference system (FIS) of order 1.

The fuzzy model is built in five steps. The first step is to define the universe of discourse of each input to the process, which corresponds to the variation between the minimum and maximum value of each input. The second step is to divide each defined universe of discourse into a number of fuzzy sets. The third step is to define the input vectors (also called input tuples) that need to be applied to the process at the experimental stage, in order to obtain the rules of the fuzzy model. In the fourth step, which is the experimental stage, a database is built from the input vectors applied to the process and the corresponding output vectors measured in the experiments. The last step is to evaluate the rules of the fuzzy model from the database of input and output vectors. This is done using an approach not used in fuzzy control: kriging interpolation.

\section{The fuzzy modeling of nonlinear processes}\label{se:Modeling}

The following subsections explain each of the five steps listed above. The notation used in this paper is introduced in each subsection.

\subsection{Defining the universe of discourse of each input}
From the knowledge available about the nonlinear process, the universe of discourse of each input is defined to make it possible to build a fuzzy model of the process. The minimal value of the $j$-th input $u_j \in \mathbb{R}$ is defined by $u_{min,j}$ and its maximal value by $u_{max,j}$. Then, the universe of discourse of the $j$-th input is $u_j \in \mathfrak{U}_j = [u_{min,j},u_{max,j}] \subset \mathbb{R}$.

The orthogonal combination of all the universes of discourses of the $m$ inputs leads to an $m$ dimension input space:
\begin{equation}
	\mathfrak{U} = \mathfrak{U}_1 \times \ldots \times \mathfrak{U}_m \in \mathbb{R}^m
\end{equation}
so the input vector $\mathbf{u} = [ u_1,\: \ldots,\: u_m ]^T \in \mathfrak{U}$. 

The universes of discourse are divided into a number of fuzzy sets.

\subsection{Dividing the universe of discourse of each input into fuzzy sets}
\begin{figure}[htbp]
	\centering
	\includegraphics[width=\textwidth]{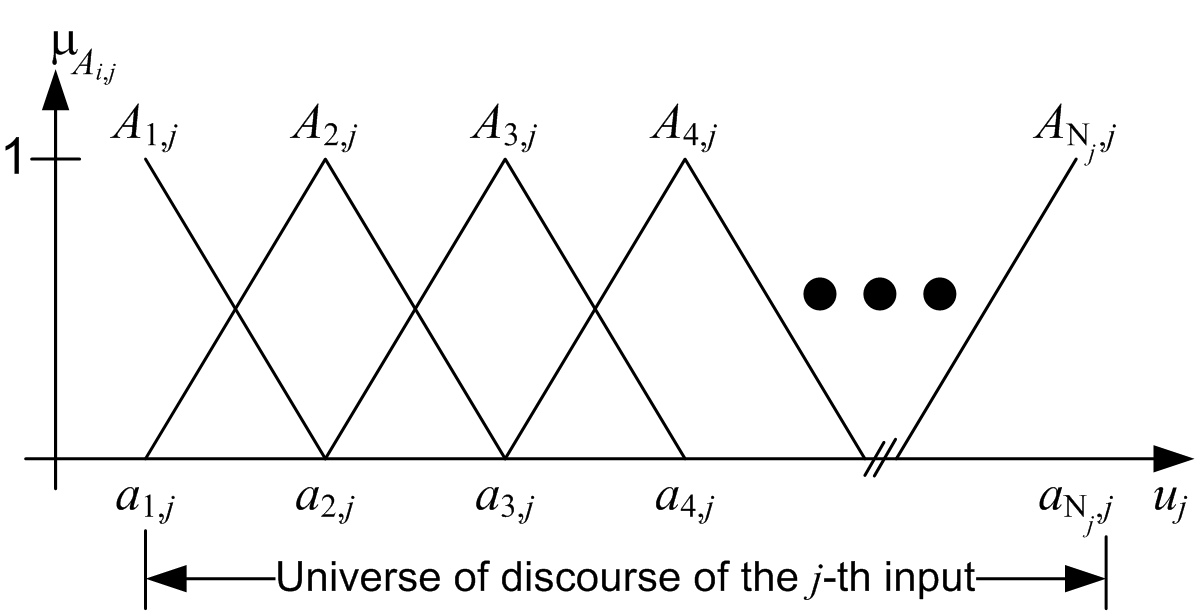}
	\caption{Membership functions of the $j$-th input fuzzy sets}
	\label{fig:universe}
\end{figure}

The universe of each input is covered by a certain number of fuzzy sets (Figure \ref{fig:universe}), each of which is identified by a linguistic term. The $i$-th fuzzy set of the $j$-th input is identified by the linguistic term $A_{i,j}$. The number of fuzzy sets defined for the $j$-th input is identified by $N_j \in \mathbb{N}$. The number of fuzzy sets has an effect on the accuracy of the fuzzy model.

For the proposed fuzzy model, the membership functions of each fuzzy set are triangular (see Figure \ref{fig:universe}). When $i \in \{ 2, \ldots, N_j-1 \}$, the membership functions are defined as follows:
\begin{equation}\label{eq:mu_mid}
	\mu_{A_{i,j}}(u_j) = 
         \begin{cases} 
	         {\mathfrak{n}_1} / {\mathfrak{m}_1} & a_{i-1,j} \leq u_j \leq a_{i,j}\\ 
		{\mathfrak{n}_2} / {\mathfrak{m}_2} & a_{i,j} < u_j \leq a_{i+1,j}\\ 
		0 & \textrm{otherwise}
         \end{cases}
\end{equation}
where $\mathfrak{n}_1 = u_j-a_{i-1,j}$, $\mathfrak{m}_1 = a_{i,j}-a_{i-1,j}$, $\mathfrak{n}_2 = a_{i+1,j}-u_j$, $\mathfrak{m}_2 = a_{i+1,j}-a_{i,j}$ and $a_{i,j}$ is the abscissa of the peak of the $A_{i,j}$ fuzzy set (Figure \ref{fig:universe}).

When $i=1$, we have the membership level of the left-most fuzzy set ($A_{1,j}$):
\begin{equation}\label{eq:mu_left}
	\mu_{A_{1,j}}(u_j) = 
         \begin{cases}
	         1 & u_j \leq u_{min,j}\\ 
		\mathfrak{n}_3 / \mathfrak{m}_3 & u_{min,j} < u_j \leq a_{2,j}\\ 
		0 & \textrm{otherwise}
         \end{cases}
\end{equation}
where $\mathfrak{n}_3 = u_j-u_{min,j}$ and $\mathfrak{m}_3 = a_{2,j}-u_{min,j}$

When $i=N_j$, we have the membership level of the right-most fuzzy set ($A_{N_j,j}$):
\begin{equation}\label{eq:mu_right}
	\mu_{A_{N_j,j}}(u_j) = 
         \begin{cases}
	         1 & u_j \geq u_{max,j}\\ 
		\mathfrak{n}_4 / \mathfrak{m}_4 & a_{N_j-1,j} \leq u_j < u_{max,j}\\ 
		0 & \textrm{otherwise}
         \end{cases}
\end{equation}
where $\mathfrak{n}_4 = u_{max,j}-u_j$ and $\mathfrak{m}_4 = u_{max,j}-a_{N_j-1,j}$.

For the proposed fuzzy model, the partitioning of the universe of discourse needs to be strict, as in \cite{Boukezzoula2001}\cite{Galichet2004}. So, for two adjacent fuzzy sets $A_{i,j}$ and $A_{i+1,j}$, the sum of their membership functions is:
\begin{equation}
	\mu_{A_{i,j}}(u_j) + \mu_{A_{i+1,j}}(u_j) = 1
\end{equation}
where $a_{i,j} \leq u_j \leq a_{i+1,j}$, for all $i \in \{1, \ldots, N_j -1 \}$ and all the $m$ inputs. Therefore, for a given input $u_j$ with a value within the universe of discourse, there are only two fuzzy sets with a value different from zero.

All the fuzzy sets have to be implemented to obtain the fuzzification component of the fuzzy model. Since the $j$-th input has $N_j$ fuzzy sets, the total number of fuzzy sets in the fuzzy model is $N_{FS} = N_1 + \ldots + N_m$.

\subsection{Defining the required input vectors necessary for experimentation on the process}

In fuzzy logic, each rule is associated with a combination of the fuzzy sets of the inputs. For example, rule  $\mathfrak{R}_{i_1,\ldots, i_m, k}$ is defined as follows:
\begin{equation}\label{eq:regle_01}
	\begin{split}
	\mathfrak{R}_{i_1,\ldots, i_m,k} :  \mathsf{If} \: u_1 \: \mathsf{is} \: A_{i_1,1} \: \mathsf{and} \: \ldots \: \mathsf{and} \: u_m \: \mathsf{is} \: A_{i_m,m}, \: \mathsf{then} \: y_k = r_{i_1,\ldots, i_m,k}
	\end{split}
\end{equation} 
In \eqref{eq:regle_01}, $k$ is the output number, and $i_j \in \{1,\ldots,N_j\}$ is used to identify the fuzzy set number considered in the rule for the $j$-th input. The consequent of rule $\mathfrak{R}_{i_1,\ldots, i_m, k}$ is expressed by $r_{i_1,\ldots, i_m, k}$. The $k$-th output $y_k$ is obtained from the rules consequent.

In this paper, we propose to create the fuzzy model rules from a database built from experiments performed on the process. The input vectors (also called input tuples) used in the experiments are defined from the fuzzy sets defined in the previous step. In the input tuple, the $j$-th input $u_j$ is such that $u_j \in \{ u_{min,j}, b_{1,j}, \ldots, b_{N_j-1,j},  u_{max,j} \}$, where the $b_{i,j}$ abscissa are:
\begin{equation}
	b_{i,j} = (a_{i,j} + a_{i+1,j})/2, \: \: i \in \{1, \ldots, N_j-1 \}
\end{equation}
and this is done for all the $m$ inputs ($j \in \{1,\ldots,m\}$).  

The input tuples used to build the database are identified by the vector $\boldsymbol{\theta}_{\{i_1,\ldots, i_m\}}$ where the indices $i_j \in \{1,\ldots,N_j + 1\}$.

Since there are $m$ inputs, the results of $2^m$ experiments are required for each rule. As the number of rules for the $j$-th input is defined as $N_j$, the total number of rules required for each output is $N = N_1 \times \ldots \times N_m$. The total number of experiments required to obtain all $N$ rules of the fuzzy model is $N_{experiments} = (N_1+1) \times \ldots \times (N_m+1)$.  The results of $2^m$ experiments are used to obtain each rule.

\subsubsection{Example}
To help explain the notation, we consider a system with two inputs. Four ($2^2$) measurements are required to obtain rule $\mathfrak{R}_{1,1,k}$, defined as:
\begin{equation}\label{eq:regle_ex01}
	\mathfrak{R}_{1,1,k} :  \mathsf{If} \: u_1 \: \mathsf{is} \: A_{1,1} \: \mathsf{and} \: u_2 \: \mathsf{is} \: A_{1,2}, \: \mathsf{then} \: y_k = r_{1,1,k}
\end{equation}
The input tuples applied to the system for those four measurements are: $\boldsymbol{\theta}_{\{1,1\}} = [u_{min,1} \: u_{min,2}]^T$, $\boldsymbol{\theta}_{\{1,2\}} = [u_{min,1} \: b_{1,2}]^T$, $\boldsymbol{\theta}_{\{2,1\}} = [b_{1,1} \: u_{min,2}]^T$ and $\boldsymbol{\theta}_{\{2,2\}} = [b_{1,1} \: b_{1,2}]^T$. Then, the four experiments are performed by setting $\mathbf{u} = \boldsymbol{\theta}_{\{i,j\}}$ with $i,j \in \{1, 2\}$. $\blacklozenge$

\subsection{Building the rules from the database}
All the experiments performed on the process provide a database linking the measured output vectors $\boldsymbol{\phi}_{\{i_1,\ldots,i_m\}} \in \mathbb{R}^m$ with each corresponding input vector (or tuple) $\boldsymbol{\theta}_{\{i_1,\ldots,i_m\}} \in \mathbb{R}^m$ (with $i_j \in \{1,\ldots,N_j+1\}$). The rules are obtained from those experiments. The  $\mathfrak{R}_{l_1,\ldots, l_m, k}$ rule of the $k$-th output is obtained from the results of  $2^m$ experiments, where the corresponding input tuples are on the following list:
\begin{equation}\label{eq:liste_teta}
	\{ \forall \: \boldsymbol{\theta}_{\{i_1,\ldots, i_m \}} \: | \: i_j \in \{l_j,l_j+1\} \} 
\end{equation} 
and the corresponding $k$-th output measurements are also on this list:
\begin{equation}\label{eq:liste_phik}
	\{ \forall \: \phi_{\{i_1,\ldots, i_m\},k} \: | \: i_j \in \{l_j,l_j+1\} \} 
\end{equation} 
The cardinality of these two lists is $2^m$.

The two lists, \eqref{eq:liste_teta} and \eqref{eq:liste_phik}, are used to find the consequent $r_{l_1,\ldots, l_m, k}$ of the rule $\mathfrak{R}_{l_1,\ldots, l_m, k}$ by using kriging interpolation. The contents of the two lists are arranged in an  $ m \times 2^m$ matrix defined as:
\begin{equation}
	\mathbf{\Theta}_{\{l_1,\ldots, l_m\}} = \left [
       \begin{array}{ccc}
       \boldsymbol{\theta}_{\{l_1,\ldots,l_m\}} & 
       \ldots & 
       \boldsymbol{\theta}_{\{l_1+1,\ldots,l_m+1\}}
       \end{array}
	\right ]^T
\end{equation} 
and the corresponding $k$-th output vector (of size $2^m$):
\begin{equation}
	\begin{split}
	& \mathbf{Y}_{\{l_1,\ldots, l_m\},k} = 
	\left [
       \begin{array}{ccc}
       \phi_{\{l_1,\ldots,l_m\},k} & 
       \ldots & 
       \phi_{\{l_1+1,\ldots,l_m+1\},k}
       \end{array}
	\right ]^T
	\end{split}
\end{equation} 

\subsubsection{Example (con't)}
Continuing the example with two inputs, the matrices used to find the rule $\mathfrak{R}_{1,1,k}$ are:
\begin{equation}
	\mathbf{\Theta}_{\{1,1\}} = \left [
       \begin{array}{cccc}
       \boldsymbol{\theta}_{\{1,1\}} & 
       \boldsymbol{\theta}_{\{1,2\}} & 
       \boldsymbol{\theta}_{\{2,1\}} &
       \boldsymbol{\theta}_{\{2,2\}}
       \end{array}
	\right ]^T
\end{equation} 
and :
\begin{equation}
	\begin{split}
	& \mathbf{Y}_{\{1,1\},k} = 
	\left [
       \begin{matrix}
       \phi_{\{1,1\},k} & 
       \phi_{\{1,2\},k} & 
       \phi_{\{2,1\},k} &
       \phi_{\{2,2\},k}
       \end{matrix}
	\right ]^T
	\end{split}
\end{equation}
$\blacklozenge$

Kringing \cite{Fortin2001} is the approach used to perform the interpolation to obtain the rules of the fuzzy model \cite{Beauchemin-Turcotte2010}. The equation used by the kriging interpolation to obtain the coefficients of the rule consequent $r_{l_1,\ldots, l_m, k}$ is the following linear system:
\begin{equation}\label{eq:kringing01}
	\begin{split}
	\left [
       \begin{matrix}
       \mathbf{I}_{2^m} & \mathbf{1}^T_{2^m} & \mathbf{\Theta}_{\{l_1,\ldots,l_m\}} \\ 
       \mathbf{1}_{2^m} & 0 & \mathbf{0}_{m} \\ 
       \mathbf{\Theta}_{\{l_1,\ldots,l_m\}}^T & \mathbf{0}^T_{m} & \mathbf{0}_{m \times m}
       \end{matrix}
	\right ] \times 
	\left [
       \begin{matrix}
       \mathbf{B}_{\{l_1,\ldots,l_m\},k} \\ 
       c_{\{l_1,\ldots,l_m\},0,k} \\ 
       \mathbf{A}_{\{l_1,\ldots,l_m\},k}
       \end{matrix}
	\right ] =  \left [
       \begin{matrix}
       \mathbf{Y}_{\{l_1,\ldots,l_m\},k} \\ 
       0 \\ 
       \mathbf{0}_{m}
       \end{matrix}
	\right ] &
	\end{split}
\end{equation} 
In \eqref{eq:kringing01}, $\mathbf{I}_{2^m} \in \mathbb{R}^{2^m \times 2^m}$ represents the identity matrix, $\mathbf{0}_{m \times m} \in \mathbb{R}^{m \times m}$ is a square matrix filled with zeros, $\mathbf{1}_{2^m} \in \mathbb{R}^{2^m}$ is a vector filled with ones and $\mathbf{0}_{m} \in \mathbb{R}^{m}$ is a vector filled with zeros. The vector $\mathbf{A}_{\{l_1,\ldots,l_m\},k} \in \mathbb{R}^{m}$ contains the list of the coefficients of the rule consequent $r_{l_1,\ldots, l_m, k}$:
\begin{equation}
\mathbf{A}_{\{l_1,\ldots, l_m\},k} = 
	\left [
       \begin{matrix}
       c_{\{l_1,\ldots, l_m\},1,k} & \ldots & c_{\{l_1,\ldots, l_m\},m,k} 
       \end{matrix}
	\right ]^T
\end{equation} 
The constant coefficient of the rule $r_{l_1,\ldots, l_m, k}$ is the variable $c_{\{l_1,\ldots, l_m\}, 0, k}$. From the coefficients contained in $\mathbf{A}_{\{l_1,\ldots,l_m\},k}$ and $c_{\{l_1,\ldots, l_m\}, 0, k}$, the rule consequent $r_{l_1,\ldots, l_m, k}$ of a TSK FIS of order 1 is defined by this affine function:
\begin{equation}
	r_{l_1,\ldots, l_m, k} = c_{\{l_1,\ldots, l_m\}, 0, k} + \sum^m_{j=1}{\left ( c_{\{l_1,\ldots, l_m\}, j, k} \times u_j \right )}
\end{equation} 

The vector $\mathbf{B}_{\{l_1,\ldots,l_m\},k} \in \mathbb{R}^{2^m}$ contains the information about the interpolation accuracy \cite{Fortin2001}. When this vector is filled with zeros, the kriging interpolation gives the exact model for the subspace corresponding to the rule $\mathfrak{R}_{i_1,\ldots,i_m,k}$, based on the $2^m$ measurements.

The coefficients of the rule $\mathfrak{R}_{i_1,\ldots,i_m,k}$ and the vector $\mathbf{B}_{\{l_1,\ldots,l_m\},k}$ are obtained by solving the linear system \eqref{eq:kringing01}.  This interpolation process is repeated $N \times m$ times, since there are $N$ rules consequent for each of the $m$ outputs.

From the interpolated rule consequent $r_{i_1,\ldots,i_m,k}$, we are able to build vectors and matrices. In fact, the rule consequents are grouped into vectors $\mathbf{R}_{\{l_1,\ldots,l_m\}} \in \mathbb{R}^{m}$ as follows:
\begin{equation}
	\mathbf{R}_{\{l_1,\ldots, l_m\}} = \left [
       \begin{array}{ccc}
       r_{l_1,\ldots,l_m,1} & \ldots & r_{l_1,\ldots,l_m,m} 
       \end{array}
	\right ]^T
\end{equation} 
and each of these vectors is defined as follows:
\begin{equation}\label{eq:rule_vector}
	\mathbf{R}_{\{l_1,\ldots, l_m\}} = \mathbf{C}_{\{l_1,\ldots, l_m\}} + \mathbf{D}_{\{l_1,\ldots, l_m\}}^T \mathbf{u}
\end{equation} 
In \eqref{eq:rule_vector}, $\mathbf{C}_{\{l_1,\ldots, l_m\}} \in \mathbb{R}^{m}$ is a vector containing the $c_{\{l_1,\ldots,l_m\},0,k}$ terms of the rules:
\begin{equation}\label{eq:vectorC}
	\mathbf{C}_{\{l_1,\ldots, l_m\}} = \left [
       \begin{array}{ccc}
       c_{\{l_1,\ldots,l_m\},0,1} & \ldots & c_{\{l_1,\ldots,l_m\},0,m} 
       \end{array}
	\right ]^T
\end{equation} 
and $\mathbf{D}_{\{l_1,\ldots, l_m\}} \in \mathbb{R}^{m \times m}$ is a matrix containing the $c_{\{l_1,\ldots,l_m\},j,k}$ terms (with $j\neq0$) of the rules:
\begin{equation}\label{eq:matrixD}
       \begin{split}
	& \mathbf{D}_{\{l_1,\ldots, l_m\}} = 
	\left [
       \begin{array}{cccc}
       c_{\{l_1,\ldots,l_m\},1,1} &  \ldots & c_{\{l_1,\ldots,l_m\},m,1}  \\
	\vdots & \ddots & \vdots  \\
       c_{\{l_1,\ldots,l_m\},1,m} & \ldots & c_{\{l_1,\ldots,l_m\},m,m}  
       \end{array}
	\right ]
	\end{split}
\end{equation} 
All the inputs of the fuzzy model are combined into a vector $\mathbf{u} = [u_1, \ldots, u_m]^T \in \mathbb{R}^{m}$.

As the kriging interpolation is linear, it is possible to obtain the matrix $\mathbf{D}$ and the vector $\mathbf{C}$ defined in \eqref{eq:matrixD} and \eqref{eq:vectorC} directly, since it is possible to rewrite \eqref{eq:kringing01} as follows:
\begin{equation}\label{eq:kringing02}
	\begin{split}
	\left [
       \begin{array}{ccc}
       \mathbf{I}_{2^m} & \mathbf{1}^T_{2^m} & \mathbf{\Theta}_{\{l_1,\ldots,l_m\}} \\ 
       \mathbf{1}_{2^m} & 0 & \mathbf{0}_{m} \\ 
       \mathbf{\Theta}_{\{l_1,\ldots,l_m\}}^T & \mathbf{0}^T_{m} & \mathbf{0}_{m \times m}
       \end{array}
	\right ] \times 
	\left [
       \begin{array}{ccc}
       \mathbf{B}_{\{l_1,\ldots,l_m\}} \\ 
       \mathbf{C}^T_{\{l_1,\ldots,l_m\}} \\
       \mathbf{D}^T_{\{l_1,\ldots,l_m\}}
       \end{array}
	\right ] = \left [
       \begin{array}{ccc}
       \mathbf{Y}_{\{l_1,\ldots,l_m\}} \\ 
       0 \\ 
       \mathbf{0}_{m}
       \end{array}
	\right ] &
	\end{split}
\end{equation} 
where the matrices $\mathbf{B}_{\{l_1,\ldots,l_m\}} \in \mathbb{R}^{m \times 2^m}$ and $\mathbf{Y}_{\{l_1,\ldots,l_m\}} \in \mathbb{R}^{m \times 2^m}$ are respectively:
\begin{equation}\label{eq:matrixB}
	\mathbf{B}_{\{l_1,\ldots, l_m\}} = \left [
       \begin{array}{cccc}
       \mathbf{B}_{\{l_1,\ldots,l_m,\},1} & \ldots & \mathbf{B}_{\{l_1,\ldots,l_m,\},m}  
       \end{array}
	\right ]
\end{equation} 
and
\begin{equation}\label{eq:matrixY}
	\mathbf{Y}_{\{l_1,\ldots, l_m\}} = \left [
       \begin{array}{cccc}
       \mathbf{Y}_{\{l_1,\ldots,l_m,\},1} & \ldots & \mathbf{Y}_{\{l_1,\ldots,l_m,\},m}  
       \end{array}
	\right ]
\end{equation} 
By solving the linear system defined in \eqref{eq:kringing02}, the vector $\mathbf{C}_{\{l_1,\ldots,l_m,\}}$ and the matrix $\mathbf{D}_{\{l_1,\ldots,l_m,\}}$ are calculated directly.

When the rule consequent are represented vectorially, the $\mathfrak{R}_{i_1,\ldots, i_m, k}$ rules, defined in  \eqref{eq:regle_01} with $k \in \{1,\ldots,m\}$, are combined into this rule:
\begin{equation}\label{eq:regle_02}
	\begin{split}
	\mathfrak{R}_{i_1,\ldots, i_m} :  \mathsf{If} \: u_1 \: \mathsf{is} \: A_{i_1,1} \: \mathsf{and} \: \ldots 
	\: \mathsf{and} \: u_m \: \mathsf{is} \: A_{i_m,m}, \: \mathsf{then} \: \mathbf{y} = \mathbf{R}_{\{i_1,\ldots, i_m\}}
	\end{split}
\end{equation} 
with $\mathbf{R}_{\{i_1,\ldots, i_m\}}$ defined in \eqref{eq:rule_vector} and the output vector $\mathbf{y} = [ y_1 \: \ldots \: y_m ]^T \in \mathbb{R}^m$.

This vectorial representation of the rule consequent simplifies the representation of the FIS rules and the inversion of the rules to obtain the inverse fuzzy model.

\subsection{The resulting fuzzy model}
By following the steps presented in the above section, a fuzzy model based on the TSK FIS of order 1 is obtained. To implement this model, $N_j$ fuzzy sets must be defined for the  $j$-th input  – see equations \eqref{eq:mu_mid} to \eqref{eq:mu_right}. This means that there are $N$ rules to define per output, resulting in  $N$ linear equations - see \eqref{eq:rule_vector}.

In the next section, the inversion of this fuzzy model is introduced.

\section{Inversion of the fuzzy model}\label{se:Inversion}

To design the fuzzy TILC algorithm, the fuzzy model of the MIMO process has to be inverted. Two steps are required to achieve this. The first step is to define the fuzzy sets of the inputs of the inverse fuzzy model from the fuzzy sets of the inputs of the fuzzy model. If this first step is feasible, the second step is to obtain the rule consequents of the inverse fuzzy model from the rule consequent of the fuzzy model.

\subsection{Preliminary assumptions}
Before explaining the inversion of the fuzzy model, we must state the following two assumptions about the TSK fuzzy model that will be inverted.

\paragraph{Assumption III-1} That the number of inputs is equal to the number of outputs, or $p = m$, and also that input $u_i$ is paired with output $y_i$ in the fuzzy model. This assumption is not very restrictive, since the input output pairing of a MIMO process can be achieved using an approach like Relative Gain Array (RGA) \cite{Ogunnaike1994}. Once this pairing has been made, the outputs are renamed, such that output $y_i$ is paired with input $u_i$.

\paragraph{Assumption III-2} That output $y_k$ of the fuzzy model, the minimum $y_{min,k}$ and the maximum $y_{max,k}$, are at opposite corners of the input space $\mathfrak{U}$, and this must be the case for all outputs ($k \in \{1,\ldots,m\}$).

\begin{figure}[!t]
	\centering
	\begin{minipage}[h]{1\linewidth}
	\center{\includegraphics[width = 0.50\linewidth]{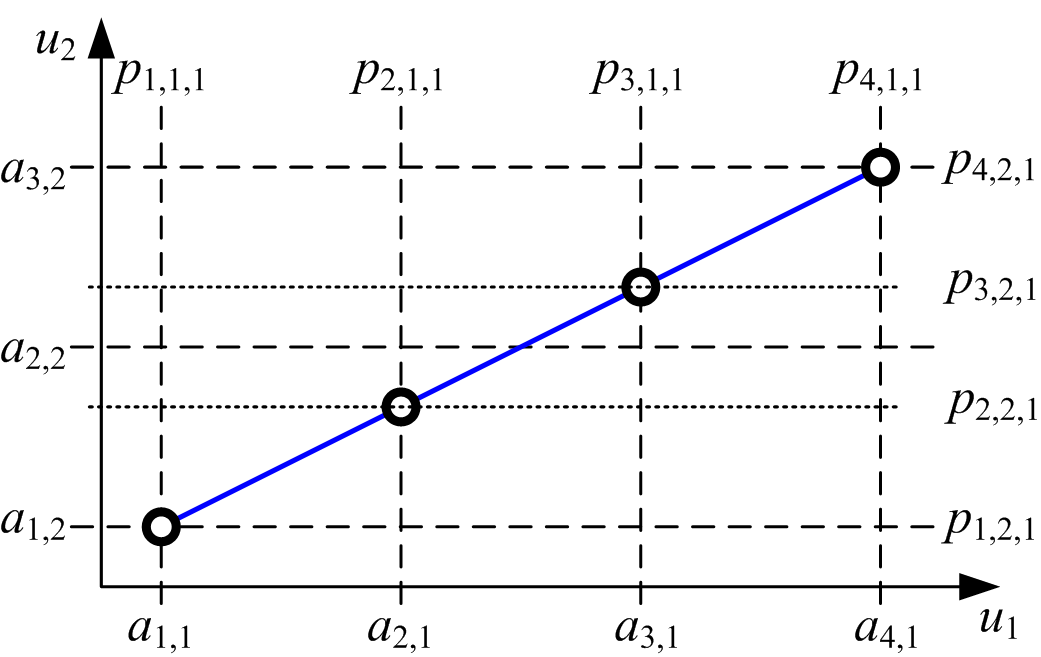} \\ a)}
	\end{minipage}
	\begin{minipage}[h]{1\linewidth}
	\center{\includegraphics[width = 0.50\linewidth]{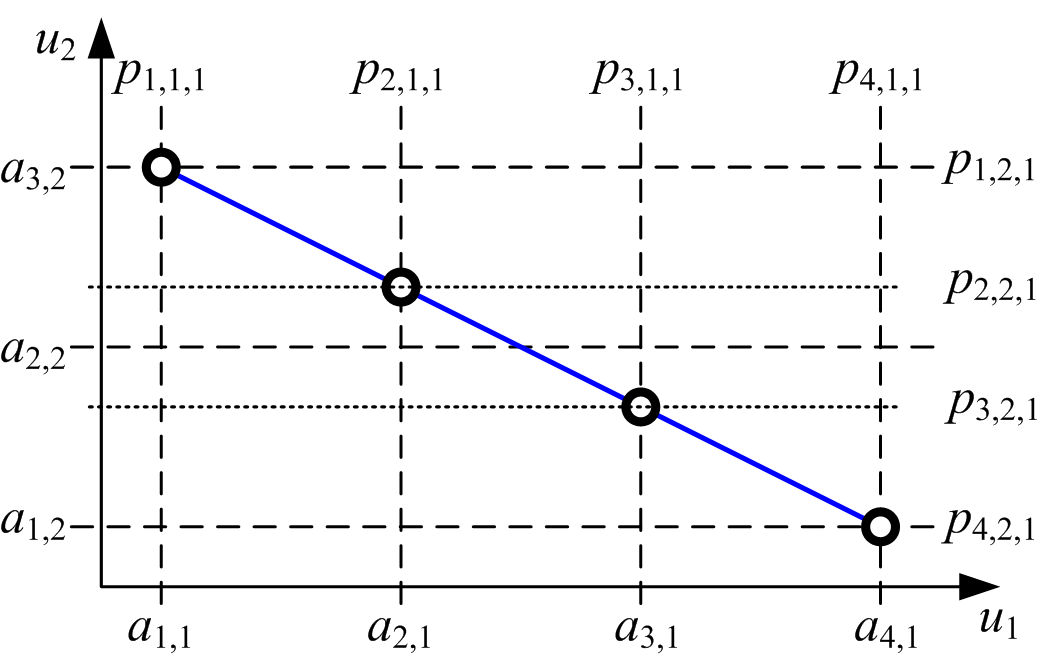} \\ b)}
	\end{minipage}	
	\caption{Diagonals along the input space $\mathfrak{U}$ of the fuzzy model for a two inputs fuzzy model}
	\label{fig:diagonal}
\end{figure}

To help in understanding the second assumption, Figure \ref{fig:diagonal} shows an example for a system having two inputs. In Figure \ref{fig:diagonal} a) the minimal and maximal values of the output are at the following input vectors: $\mathbf{u} = [a_{1,1} \: \: a_{1,2}]^T$ and $\mathbf{u} = [a_{N_1,1} \: \: a_{N_2,2}]^T = [a_{4,1} \: \: a_{3,2}]^T$. Figure \ref{fig:diagonal} b) the minimal and maximal values of the output are at $\mathbf{u} = [a_{1,1} \: \: a_{N_2,2}]^T = [a_{1,1} \: \: a_{3,2}]^T$ and $\mathbf{u} = [a_{N_1,1} \: \: a_{1,2}]^T = [a_{4,1} \: \: a_{1,2}]^T$.

If neither of the cases presented in Assumption III-2 applies, the approach described in the following subsection for performing the inversion of the fuzzy model cannot be used.

\subsection{Obtaining the fuzzy sets of the inverse fuzzy model}
To obtain the fuzzy model of a process, fuzzy sets are defined for the $m$ inputs of the process. In the inverse fuzzy model, the fuzzy sets are built from those fuzzy sets. Since we have assumed that the number of inputs and outputs of the process is the same, the number of fuzzy sets in the inverse fuzzy model is equal to those in the fuzzy model.

The inputs to the inverse fuzzy model are defined by the variable $\tilde{y}_j$, and they are grouped in the vector $\tilde{\mathbf{y}} = [\tilde{y}_1 \: \ldots \: \tilde{y}_m]^T \in \mathbb{R}^m$. The linguistic term of the $i$-th fuzzy set of $k$-input $\tilde{y}_k$ of the inverse fuzzy model is identified by $\mathbb{A}_{i,k}$. The coordinates of the peak of each of those fuzzy sets are identified by  $\mathfrak{a}_{i,k}$. Each of the $\mathfrak{a}_{i,k}$ coordinates is calculated from the $a_{i,j}$ coordinates from the fuzzy model in a two steps process.

To explain this process, let us consider a two input, two output fuzzy model. Figure \ref{fig:diagonal} shows the universe of discourse of both the inputs and the coordinates ($a_{i,j}$) of the peaks of the fuzzy sets of both the inputs. The output $y_1$ of this fuzzy model is assumed to be minimal when input $u_1 = a_{1,1}$ and input $u_2 = a_{1,2}$, and maximal when $u_1 = a_{4,1}$ and $u_2 = a_{3,2}$.

Each of the peaks $\mathfrak{a}_{i,1}$ of the fuzzy sets $\mathbb{A}_{i,1}$ of the input $\tilde{y}_1$ of the inverse fuzzy model is obtained as follows:

First, the coordinates $p_{i,j,k}$, where the fuzzy sets of the input $u_1$  have their peaks, are calculated. Then, for this input, the coordinates $p_{i,1,1}$ are the coordinates $a_{i,1}$ directly, and so $p_{i,1,1} = a_{i,1}$. For the other input ($u_2$), the coordinates to consider are identified as $p_{i,2,1}$, and are calculated by the equations below (when $j \neq k$).

For case A in Figure \ref{fig:diagonal}:
\begin{equation}\label{eq:pijk_caseA}
	\begin{split}
	p_{i,j,k} &= \frac{(a_{N_j,j} - a_{1,j})}{(a_{N_k,k} - a_{1,k})}(a_{i,k} - a_{1,k}) + a_{1,j} \\ &= \frac{(u_{max,j} - u_{min,j})}{(u_{max,k} - u_{min,k})}(a_{i,k} - u_{min,k}) + u_{min,j}
	\end{split}
\end{equation} 
and for case B in Figure \ref{fig:diagonal}:
\begin{equation}\label{eq:pijk_caseB}
	\begin{split}
	p_{i,j,k} &= \frac{(a_{N_j,j} - a_{1,j})}{(a_{N_k,k} - a_{1,k})}(a_{N_k,k} - a_{i,k}) + a_{1,j} \\ &= \frac{(u_{max,j} - u_{min,j})}{(u_{max,k} - u_{min,k})}(u_{max,k} - a_{i,k}) + u_{min,j}
	\end{split}
\end{equation}

In both cases, when $j=k$ (the number of inputs and outputs is the same), the equation to use is $p_{i,k,k} = a_{i,k}$.

Finally, the coordinates $\mathfrak{a}_{i,k}$ are obtained directly from the evaluation of the fuzzy model with the input vector built from those $p_{i,j,k}$ values. So, $\mathfrak{a}_{i,k}$ is directly, the $k$-th output value $y_k$ of the fuzzy model with the vector $[p_{i,1,k} \: \ldots$ $\: p_{i,m,k}]^T$ applied as the input vector.

In this paper, we consider the following two cases. If $\mathfrak{a}_{i,k}$ are not increasing, then  $\mathfrak{a}_{1,k} \leq \mathfrak{a}_{2,k} \leq \ldots \leq \mathfrak{a}_{N_k,k}$. If $\mathfrak{a}_{i,k}$ are not decreasing, then $\mathfrak{a}_{1,k} \geq \mathfrak{a}_{2,k} \geq \ldots \geq \mathfrak{a}_{N_k,k}$. We check for all $k$ (or all input output pairs), if either case applies. If neither of these two cases applies, the approach presented in this paper cannot be used. In this situation, it is not possible to design a TILC algorithm based on the inverse of the fuzzy model.

\subsection{Obtaining the rules of the inverse fuzzy model}
The rules of the fuzzy inverse model are defined in a similar way to the rules of the fuzzy model \eqref{eq:regle_01}:
\begin{equation}\label{eq:inv_regle_01}
	\begin{split}
	\mathcal{R}_{i_1,\ldots, i_m} :  \mathsf{If} \: \tilde{y}_1 \: \mathsf{is} \: \mathbb{A}_{i_1,1} \: \mathsf{and} \: \ldots 
	\: \mathsf{and} \: \tilde{y}_m \: \mathsf{is} \: \mathbb{A}_{i_m,m}, \: \mathsf{then} \: \tilde{u}_k = \mathfrak{r}_{i_1,\ldots, i_m, k}
	\end{split}
\end{equation} 
By grouping the rules as was done in equation \eqref{eq:regle_02} in the previous section, we can write:
\begin{equation}\label{eq:inv_regle_02}
	\begin{split}
	\mathcal{R}_{i_1,\ldots, i_m} :  \mathsf{If} \: \tilde{y}_1 \: \mathsf{is} \: \mathbb{A}_{i_1,1} \: \mathsf{and} \: \ldots 
	\: \mathsf{and} \: \tilde{y}_m \: \mathsf{is} \: \mathbb{A}_{i_m,m}, \: \mathsf{then} \: \mathbf{\tilde{U}} = \mathsf{R}_{\{i_1,\ldots, i_m\}}
	\end{split}
\end{equation} 
where $\mathbf{\tilde{u}}=[\tilde{u}_1 \: \ldots \: \tilde{u}_m]^T$ is the output vector of the inverse fuzzy model. Since the rules were expressed by the linear equations shown in  \eqref{eq:rule_vector}, the inversion of the rules is calculated as follows:
\begin{equation}\label{eq:invrule_vector}
	\mathsf{R}_{\{l_1,\ldots, l_m\}} = \mathsf{C}_{\{l_1,\ldots, l_m\}} + \mathsf{D}_{\{l_1,\ldots, l_m\}} \mathbf{\tilde{y}}
\end{equation} 
In \eqref{eq:invrule_vector}, the matrix $\mathsf{D}_{\{l_1,\ldots, l_m\}} \in \mathbb{R}^{m \times m}$ is calculated as follows:
\begin{equation}\label{eq:invD}
	\mathsf{D}_{\{l_1,\ldots, l_m\}} = \mathbf{D}^{-1}_{\{l_1,\ldots, l_m\}}
\end{equation}
and so the vector $\mathsf{C}_{\{l_1,\ldots, l_m\}} \in \mathbb{R}^m$ is the following:
\begin{equation}\label{eq:invC}
	\mathsf{C}_{\{l_1,\ldots, l_m\}} = -\mathbf{D}^{-1}_{\{l_1,\ldots, l_m\}}\mathbf{C}_{\{l_1,\ldots, l_m\}}
\end{equation}

As we can see, the rules for the inverse fuzzy model are easily calculated using linear algebra, i.e. by inverting the equation of the rules of the fuzzy model of the process.

Having addressed the decoupling component of the TILC algorithm, in the next section, we now cover the filter component of the TILC approach.

\section{The filter component of the TILC}\label{se:Filter}

The IMC-based fuzzy TILC is built by combining two components: the inverse fuzzy model used for decoupling, and the filter (see Figure \ref{fig:Proposed_TILC}).  

We have shown above how the model is obtained and inverted. The fuzzy model is an approximation of the real process, and this process can have parametric variations and external perturbations. The filter component of the TILC can compensate for model inaccuracy and perturbations \cite{Morari1989}.

\subsection{Crisp version of the filter}
For the IMC approach, an exponential filter is usually proposed \cite{Morari1989}. For the $i$-th loop, the exponential filter is defined by:
\begin{equation}\label{eq:kth_crisp_filter}	
\hat{Q}_i(z)=\frac{z(1-\alpha_i)}{z-1}
\end{equation}
where the filter parameter $0 \leq \alpha_i <1$ is adjusted to shape the behavior of the closed loop response. For the TILC, those $\alpha_i$ parameters have an effect on the convergence speed and the robustness of the closed loop system.

For a system with a MIMO IMC for $m$ loops, the MIMO version of the exponential filter is as follows:
\begin{equation}\label{eq:m_crisp_filter}
	\mathbf{\hat{Q}}(z)=\frac{z}{z-1} \left[
	\begin{matrix}
		1-\alpha_1 & 0 & \ldots & 0 \\
		0 & 1-\alpha_2 & \ldots & 0\\
		\vdots & \vdots & \ddots & \vdots\\
		0 & 0 & \ldots & 1-\alpha_m
	\end{matrix} \right]
\end{equation}

\subsection{Fuzzy version of the filter}

\begin{figure*}[t]
	\centering
	\includegraphics[width=\textwidth]{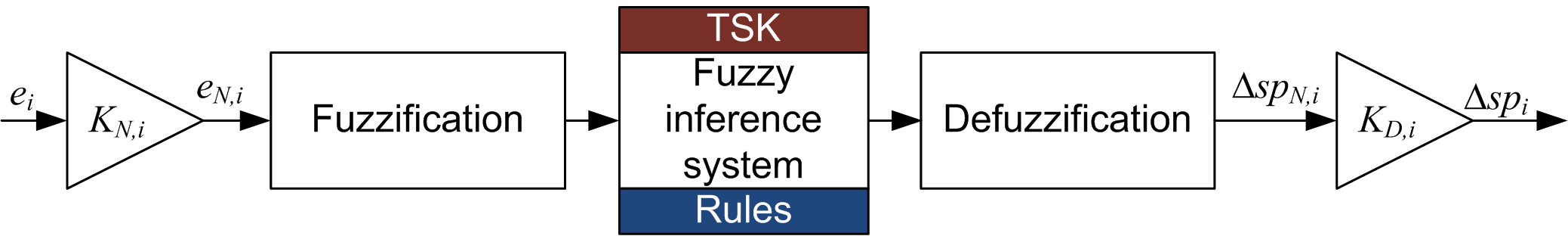}
	\caption{Block diagram of a SISO fuzzy filter}
	\label{fig:fuzzy_filter}
\end{figure*}

The fuzzy version of the IMC filter is nonlinear, since the behavior of the filter is adjusted to the temperature error \cite{Beauchemin-Turcotte2010}. 

If the temperature at the surface of the plastic sheet is too high, the heater temperature setpoints have to be decreased relatively quickly. When the temperature at the surface of the plastic sheet is too low, the heater temperature setpoints have to be increased, but at a slower speed to ensure a monotonic convergence. The purpose of the nonlinear filter is to reduce the risk of high temperature setpoints, which can damage the thermoforming oven. 

The proposed fuzzy filter (Figure \ref{fig:fuzzy_filter}) is a based on a SISO TSK FIS with constant consequents \cite{Beauchemin-Turcotte2010}. The consequent of a given rule is a constant value. The $i$-th surface temperature error $e_i$ is normalized to $e_{N,i}$ with a normalising gain $K_{N,i} \in \mathbb{R}$. The fuzzification component receives this normalized value and fuzzifies it with five fuzzy sets, and corresponding to those fives fuzzy sets there are five rules and their corresponding consequent:
\begin{equation}\label{eq:regle_filter01}
	\mathfrak{R}_{f1} :  \mathsf{If} \: e_N \: \mathsf{is} \: A_{NB} \: \mathsf{then} \: \Delta u_N = 0.6
\end{equation} 
\begin{equation}\label{eq:regle_filter02}
	\mathfrak{R}_{f2} :  \mathsf{If} \: e_N \: \mathsf{is} \: A_{NS} \: \mathsf{then} \: \Delta u_N = 0.25
\end{equation} 
\begin{equation}\label{eq:regle_filter03}
	\mathfrak{R}_{f3} :  \mathsf{If} \: e_N \: \mathsf{is} \: A_{ZR} \: \mathsf{then} \: \Delta u_N = 0.0
\end{equation} 
\begin{equation}\label{eq:regle_filter04}
	\mathfrak{R}_{f4} :  \mathsf{If} \: e_N \: \mathsf{is} \: A_{PS} \: \mathsf{then} \: \Delta u_N = -0.5
\end{equation} 
\begin{equation}\label{eq:regle_filter05}
	\mathfrak{R}_{f5} :  \mathsf{If} \: e_N \: \mathsf{is} \: A_{PB} \: \mathsf{then} \: \Delta u_N = -1.0
\end{equation} 

We can see that, as required, the temperature setpoint change $\Delta sp_{N,i}$ is greater when the sheet temperature is higher than the desired one.

The rule consequent are defuzzified, and the resulting value is defuzzified with the gain $K_{D,i} \in \mathbb{R}$ and denormalized to obtain the $i$-th output value $\Delta sp_i$. Then, the output of the filter is, finally: $sp_i[k+1] = sp_i[k] + \Delta sp_i[k]$.  This output is sent to the fuzzy inverse model to generate the heater temperature setpoints. 

The initial value $sp_i[0]$ (for cycle $k=0$) of the fuzzy filter is equal to the desired terminal temperature $y_{d,i}$. These initial values ($\forall i \in \{1,\ldots,m\}$) will be used to calculate the initial guess for the heater temperature setpoints.  

\section{Application to a thermoforming machine}\label{se:Simulation}

The fuzzy TILC design approach presented in this paper is now applied to a nonlinear model of a thermoforming machine. This model, presented briefly in the Appendix, is based on the AAA thermoforming machine \cite{Gauthier2008}. The simulated thermoforming machine has six inputs (heaters) and six outputs (infra-red sensors).

\begin{figure}[htbp]
	\centering
	\includegraphics[width = 0.50 \linewidth]{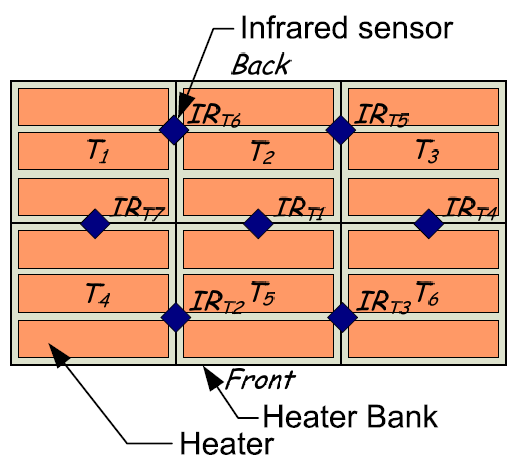}
	\caption{Location of infrared sensors and heaters (bottom sensors and heaters share the same location) \cite{Gauthier2008}}
	\label{fig:heater}
\end{figure}

The six heaters are obtained by grouping the heaters bank (see Figure \ref{fig:heater}) as follows: $T_1$-$T_4$, $T_2$-$T_5$, $T_3$-$T_6$, $B_1$-$B_4$, $B_2$-$B_5$, and  $B_3$-$B_6$ ($T$ identifies a top heater, and $B$ identifies a bottom heater). Furthermore, the system input vector $\mathbf{u} \in \mathbb{R}^6$ is defined as follows:
\begin{equation}
	\mathbf{u} = \left[
	\begin{matrix}
	u_1 \\ u_2 \\ u_3 \\ u_4 \\ u_5 \\ u_6
	\end{matrix} \right] 
	= \left[
	\begin{matrix}
	T_2-T_5 \\ T_1-T_4 \\ T_3-T_6 \\ B_2-B_5 \\ B_1-B_4 \\ B_3-B_6 
	\end{matrix} \right] 
\end{equation}

The six infrared (IR) sensors selected are (see Figure \ref{fig:heater}): $IR_{T1}$, $IR_{T2}$, $IR_{T5}$, $IR_{B1}$, $IR_{B2}$, and $IR_{B5}$ (again, $T$ identifies a top IR sensor, $B$  identifies a bottom IR sensor).  So, the model output vector $\mathbf{y} \in \mathbb{R}^6$ is defined as follows:
\begin{equation}
	\mathbf{y} = \left[
	\begin{matrix}
	y_1 \\ y_2 \\ y_3 \\ y_4 \\ y_5 \\ y_6
	\end{matrix} \right] 
	= \left[
	\begin{matrix}
	IR_{T1} \\ IR_{T2} \\ IR_{T5} \\ IR_{B1} \\ IR_{B2} \\ IR_{B5}
	\end{matrix} \right] 
\end{equation}

\subsection{Simulation setup}
The objective of the controllers is to maintain the terminal surface temperature error within a 5$^\circ$C range. However, a 10$^\circ$C error is considered acceptable for quality control.

The thermoforming model is used as benchmark to test the following three TILC algorithms:
\begin{itemize}
\item[$\bullet$] Crisp 1st order TILC – used as reference for comparison purposes;
\item[$\bullet$] Ideal fuzzy TILC (i-f-TILC) – built from non noisy data;
\item[$\bullet$] Noisy fuzzy TILC (n-f-TILC) – built from noisy data. The noise is assumed to be Gaussian, with variance $\sigma_{noise}^2$ on all outputs. For the simulations, the measurement noise of each channel is considered to be independent from that of the others. Since the data contain noise, thirty n-f-TILC were designed and tested to determine the effect of noise on the design.
\end{itemize}
All the fuzzy TILC were derived from a fuzzy model obtained from the nominal thermoforming model. Three cases were analyzed:
\begin{itemize}
\item[$\bullet$] Case A: target temperature fixed to  $\mathbf{y}_d = [160, 150, 150, 160, 150, 150]^T \: ^\circ$C;  
\item[$\bullet$] Case B: target temperature fixed to  $\mathbf{y}_d = [140, 140, 140, 140, 140, 140]^T \: ^\circ$C (uniform surface temperature);  
\item[$\bullet$] Case C: target temperature fixed to  $\mathbf{y}_d = [160, 150, 150, 160, 150, 150]^T \: ^\circ$C and $\sigma_{noise}$ is fixed to 1, 2, 3, 4 and 5 $^\circ$C.
\end{itemize}

\begin{figure*}[!t]
	\centering
	\includegraphics[width = 1 \linewidth]{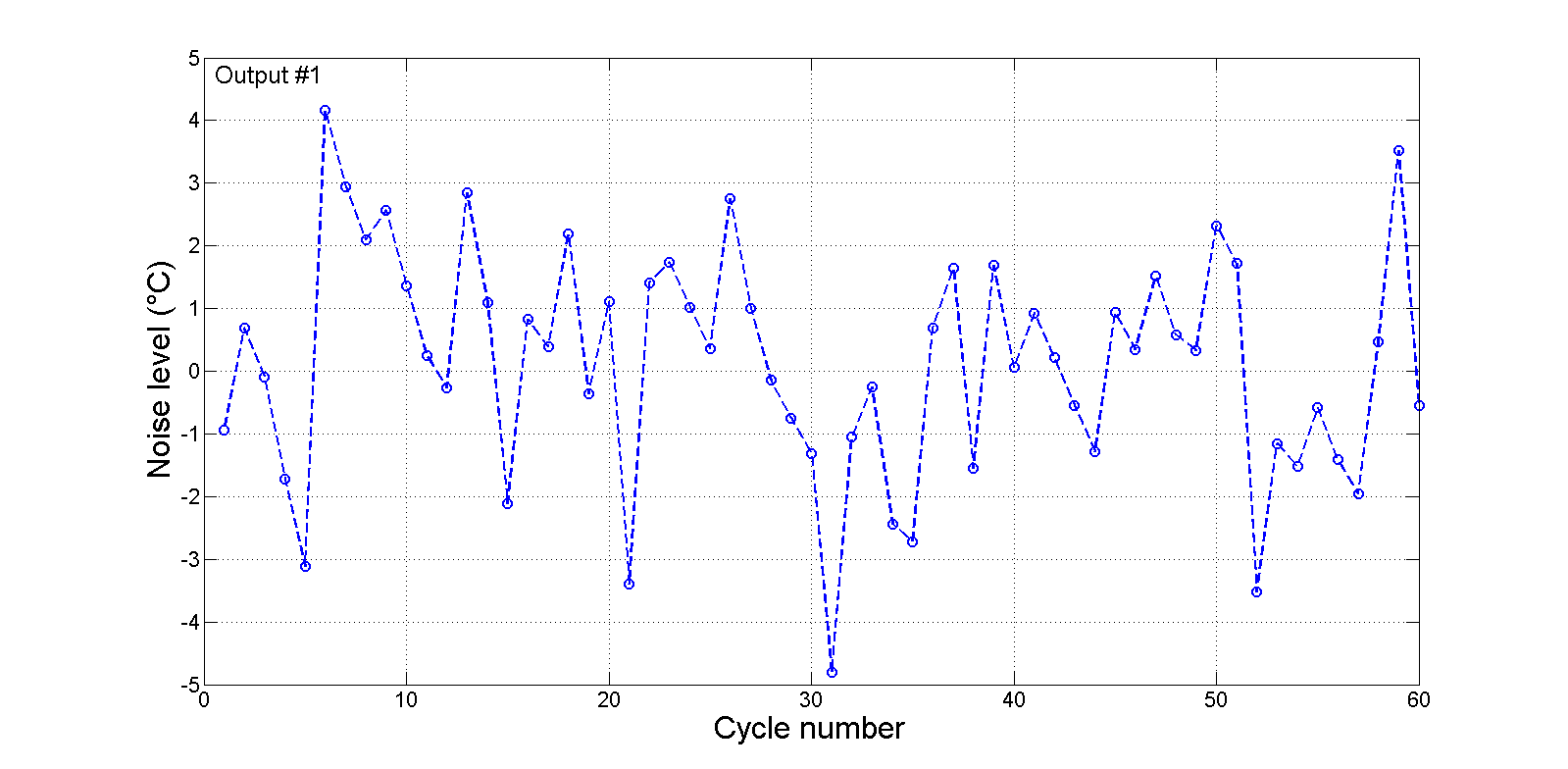}
	\caption{Gaussian noise added to output $y_1$ ($^\circ$C) (when $\sigma_{noise} = 2^\circ$C).}
	\label{fig:T_error_01}
\end{figure*}

In cases A and B, the simulations were performed on both nominal and disturbed thermoforming models. The parameters of the two models are shown in Table \ref{ta:table_parametres}. Both models were tested with and without noise ($\sigma_{noise} = 2 ^\circ$C).  To enable comparison of the three TILC algorithms, the measurement noise vectors were exactly the same for all simulations. Figure \ref{fig:T_error_01} shows the noise added to output $y_1$ in all simulations with noise. The other noise signals added to the outputs $y_2$ to $y_6$ were fairly similar, and were generated with Matlab\textsuperscript{\circledR}'s "randn" function.

\begin{table}[htbp]
	\caption{Parameters used in the simulations}
	\centering
\begin{tabular}{|c||c|c|c|}
		\hline
		Parameters & Units & Nominal value & Disturbed value \\
		\hline
		\hline
		Density & kg/m$^3$ & 950 & 1045 \\
		\hline
		Specific heat & J/(kg K) & 1838 & 2022 \\
		\hline
		Effective emissivity & & 0.45 & 0.495 \\
		\hline
		Absorptivity & m$^{-1}$ & 300 & 350 \\
		\hline
		Heat conduction & W/(m K) & 0.4 & 0.3 \\
		\hline
		Convection factor & W/(m$^2$ K) & 6 & 10 \\
		\hline
	\end{tabular}
	\label{ta:table_parametres}
\end{table}

We considered the ambient temperature (in the oven) to be constant in the simulations: $T_{amb} = 125 ^\circ$C, or to be subject to drift during the day, so that $T_{amb} = 125 ^\circ$C $+ 20^\circ$C $\sin(0.0175 k)$ (where $k$ is the cycle number). The drift reflects a situation occurring in the industry that forces the operator of the thermoforming machine to retune the heater setpoints to maintain the quality of the parts produced.

Case C is included to determine the effect of different noise levels on the n-f-TILC. Since the fuzzy controller design is affected by noise, we need to check whether or not the noise level has an adverse effect on the design. The results are presented for the worst case, i.e. the disturbed thermoforming model with both noise $\sigma_{noise}$ and drift. The n-f-TILC has been designed with the same level of noise as used in the simulation. The Kruskal-Wallis statistical test \cite{Rakotomalada2008} is used to check, with a confidence level of 95\% ($\alpha=0.05$), the similarity of the results. In this test, the $\mathcal{H}_0$ hypothesis is that all fuzzy TILCs behave similarly from cycle 10 to 60, and the $\mathcal{H}_1$ hypothesis is that at least one fuzzy TILC behaves differently from the others. The threshold level is $\chi^2_{0.05;30} = 43.77$ for the 31 fuzzy TILCs.

The simulations were performed over 60 cycles ($k \in \{1,\ldots,60\}$) and each cycle corresponded to 12 mm plastic sheets being heated for 300 seconds (5 minutes).

\subsection{Parameters for obtaining the inverse fuzzy model}
The fuzzy models of the thermoforming machine were developed by defining a universe of discourse for each input, where $u_{min,k} = 300^\circ$C and $u_{max,k} = 450^\circ$C (with $k \in \{ 1, \ldots , 6 \}$). Each universe of discourse is divided among three fuzzy sets with the peaks located at: $a_{1,k} = 300^\circ$C, $a_{2,k} = 375^\circ$C and $a_{1,k} = 450^\circ$C (with $k \in \{ 1, \ldots , 6 \}$) respectively.

There are $3^6=729$ rules for obtaining each output. This requires $4^6=4096$ experiments (simulations, here) to be performed on the thermoforming machine model.

The inverse of the fuzzy model involves evaluating the universe of discourse of all the outputs. The lower limits of those universes of discourse are obtained when all six heating elements are at the lowest setpoint: 300$^\circ$C. Without measurement noise (and nominal parameters), those lower limits are: $y_{min,k} = 105.06^\circ$C (for $k={1,3,4,6}$) and $y_{min,k} = 117.03^\circ$C (for $k={2,5}$). The higher limits are obtained when all the heating elements are at the highest setpoint: 450$^\circ$C. Again, with no measurement noise, these limits are: $y_{max,k} = 203.89^\circ$C (for $k={1,3,4,6}$) and $y_{max,k} = 233.30^\circ$C (for $k={2,5}$). So, as assumed, the minimum and maximum output temperatures are at opposite sides of the input space $\mathfrak{U}$. Since the inputs of the fuzzy model were fuzzified with three fuzzy sets each, the same applies to the inputs of the inverse fuzzy model.  

All the fuzzy filters of the fuzzy TILC have normalization gains $K_{N,i} = 0.25$ and denormalization gains $K_{D,i} = 1.00$ for $i \in \{1,\ldots,6\}$.

The two fuzzy TILCs are compared with a so-called crisp TILC based on a linear IMC approach, with an exponential filter \cite{Gauthier2008}\cite{Morari1989} where the parameters are $\alpha_i = 0.2701 \: \forall i\in \{1,\ldots,6\}$.

\subsection{Simulation results}
The simulation results are presented below to compare the behavior of the three TILCs in different situations.

\subsubsection{Case A}
The first set of simulations has a target temperature vector $\mathbf{y}_d = [160, 150,$ $150,$ $160, 150, $ $150]^T \: ^\circ$C. The heater temperature setpoints used for the crisp TILC at cycle $k=1$ are arbitrarily fixed to 350$^\circ$C for all the heaters. This value is a bad guess, as shown by the surface temperature error at cycle $k=1$ (or $||e[1]||_\infty$) displayed in last column of Tables \ref{ta:no_noise}, \ref{ta:onlynoise_2}, \ref{ta:onlydrift} and \ref{ta:noisy}. It is to be hoped that the crisp TILC algorithm will reduce this error to a lower level as the cycle number $k$ increases. In the ideal case (no noise and no drift), the error converges to zero.

For the two fuzzy TILCs, the initial heater temperature setpoints are obtained from the inverse fuzzy model. With the i-f-TILC, the first heater temperature setpoints are in the vector $\mathbf{sp} = [399.97, 353.22,$ $403.16, 399.97, 353.22,$ $403.16]^T \: ^\circ$C. As shown in Tables \ref{ta:no_noise} to \ref{ta:noisy}, the surface temperature error obtained in the first cycle or $||e[1]||_\infty$ is lower than the error obtained with the crisp TILC. Since 30 n-f-TILCs were tested, the values displayed are the mean of the 30 n-f-TILC guesses. The differences between the initial setpoint evaluated by the i-f-TILCs and those evaluated by the 30 n-f-TILCs seem to be negligible. 

TILC algorithms must converge to a setpoint vector $\mathbf{sp}$, such that the surface temperature error  $||e[k]||_\infty$ falls to 0. When there is no noise, all TILC errors converge to 0. But, when there is noise, there is an error (see $\mu_e$ and $\sigma_e$ in Tables \ref{ta:onlynoise_2} to \ref{ta:noisy}). $\mu_e$ is the mean error during cycles 10 to 60, and $\sigma_e$ is the standard deviation associated with this mean.

\begin{table}[htbp]
\center
\caption{Simulation results (Case A) --- without noise and drift}
\begin{tabular}{|l|l|r|r|r|r|r|}
\hline
TILC algorithm & System & $||e[1]||_\infty$ ($^\circ$C) \\
\hline
\multirow{2}{*}{Crisp TILC} &
Disturbed & 20.7607 \\
& Nominal & 18.8756 \\
\hline
\multirow{2}{*}{i-f-TILC} &
Disturbed & 5.5493 \\
 & Nominal & 1.0671 \\
\hline
n-f-TILC &
Disturbed & 5.8967 \\
(mean of 30 controllers)  & Nominal & 1.5671 \\
\hline
\end{tabular}
\label{ta:no_noise}
\end{table}

As shown in Table \ref{ta:onlynoise_2}, when there is noise ($\sigma_{noise} = 2^\circ$C), both fuzzy TILCs outperform the crisp TILC by nearly 1.25$^\circ$C, and the standard deviation $\sigma_e$ of the surface temperature error is divided by 1.5.

\begin{table}[htbp]
\center
\caption{Simulation results (Case A) --- with noise ($\mu_e$ and $\sigma_e$ for $k \in \{10,\ldots,60\}$)}
\begin{tabular}{|l|l|r|r|r|r|r|}
\hline
TILC algorithm & System & $\mu_e$ ($^\circ$C) & $\sigma_e$ ($^\circ$C) & $||e[1]||_\infty$ ($^\circ$C) \\
\hline
\multirow{2}{*}{Crisp TILC} &
Disturbed & 5.1039 & 1.6734 & 26.6153 \\
& Nominal & 4.9209 & 1.5830 & 24.7302 \\
\hline
\multirow{2}{*}{i-f-TILC} &
Disturbed & 3.8572 & 1.0799 & 10.5062 \\
 & Nominal & 3.6603 & 1.0521 & 6.9217 \\
\hline
n-f-TILC &
Disturbed & 3.8653 & 1.0887 & 10.4140 \\
(mean of 30 controllers) & Nominal & 3.6643 & 1.0539 & 6.8224 \\
\hline
\end{tabular}
\label{ta:onlynoise_2}
\end{table}

\begin{table}[htbp]
\center
\caption{Simulation results (Case A) --- with drift ($\mu_e$ and $\sigma_e$ for $k \in \{10,\ldots,60\}$)}
\begin{tabular}{|l|l|r|r|r|r|r|}
\hline
TILC algorithm & System & $\mu_e$ ($^\circ$C) & $\sigma_e$ ($^\circ$C) & $||e[1]||_\infty$ ($^\circ$C) \\
\hline
\multirow{2}{*}{Crisp TILC} &
Disturbed & 0.0549 & 0.0201 & 20.7607 \\
& Nominal & 0.0417 & 0.0270 & 18.8756 \\
\hline
\multirow{2}{*}{i-f-TILC} &
Disturbed & 0.3243 & 0.2531 &  5.5493 \\
 & Nominal & 0.1813 & 0.0202 & 1.0671 \\
\hline
n-f-TILC &
Disturbed & 0.3585 & 0.3194 & 5.8967 \\
(mean of 30 controller) & Nominal & 0.1894 & 0.0212 & 1.5671 \\
\hline
\end{tabular}
\label{ta:onlydrift}
\end{table}

When the ambient temperature is subject to drift and no measurement noise, there is a small error (Table \ref{ta:onlydrift}). This slowly varying error remains low, and the crisp TILC gives a lower steady state error than either of the fuzzy TILCs.

\begin{table}[htbp]
\center
\caption{Simulation results (Case A) --- with noise and drift ($\mu_e$ and $\sigma_e$ for $k \in \{10,\ldots,60\}$)}
\begin{tabular}{|l|l|r|r|r|r|r|}
\hline
TILC algorithm & System & $\mu_e$ ($^\circ$C) & $\sigma_e$ ($^\circ$C) & $||e[1]||_\infty$ ($^\circ$C) \\
\hline
\multirow{2}{*}{Crisp TILC} &
Disturbed & 5.0559 & 1.6482 & 26.6153 \\
& Nominal & 4.9005 & 1.5933 & 24.7302 \\
\hline
\multirow{2}{*}{i-f-TILC} &
Disturbed & 3.6883 & 0.9990 & 10.5150 \\
 & Nominal & 3.5901 & 1.0077 & 6.9217 \\
\hline
n-f-TILC &
Disturbed & 3.6950 & 1.0122 & 10.4140 \\
(mean of 30 controllers) & Nominal & 3.5938 & 1.0059 & 6.8224 \\
\hline
\end{tabular}
\label{ta:noisy}
\end{table}

\begin{figure*}[!t]
	\centering
	\includegraphics[width = 1 \linewidth]{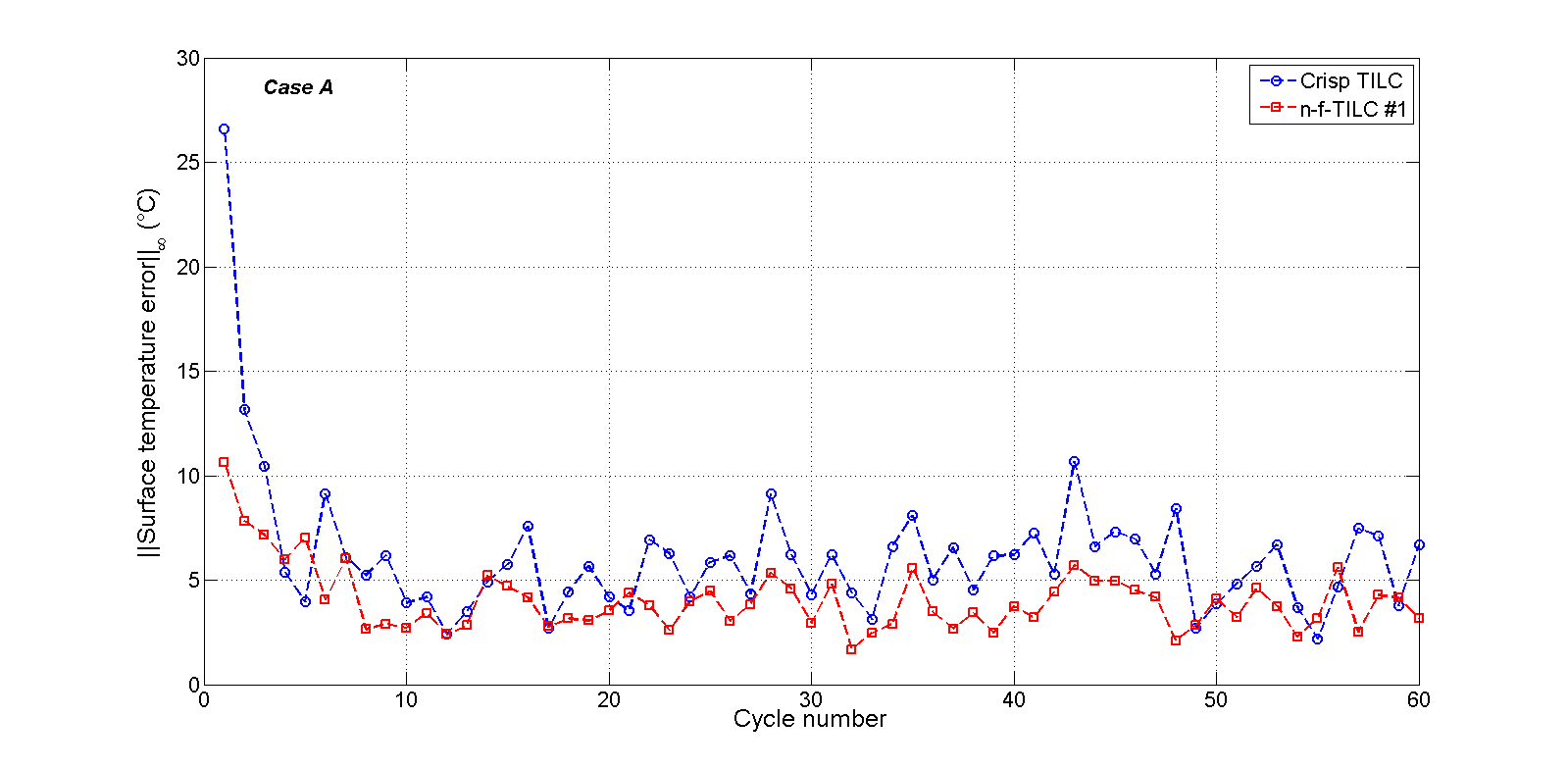}
	\caption{Infinity norm of the surface temperature error in the case of a disturbed system with noise ($\sigma_{noise}=2^\circ$C) and drift.}
	\label{fig:T_error_02}
\end{figure*}

Table \ref{ta:noisy} shows the results when we have both noise and drift. The results look similar to the results in Table \ref{ta:onlynoise_2}. The drift effect seems relatively small compared to the noise effect. Figure \ref{fig:T_error_02} shows the worst case, when the system is subjected to noise, drift, and parameter disturbance. This figure shows that once the TILC algorithms have converged, all the surface temperature errors remain under 12$^\circ$C. Furthermore, the \#1 n-f-TILC's error remains below 6$^\circ$C after cycle 7.

\begin{figure*}[!t]
	\centering
	\includegraphics[width=\textwidth]{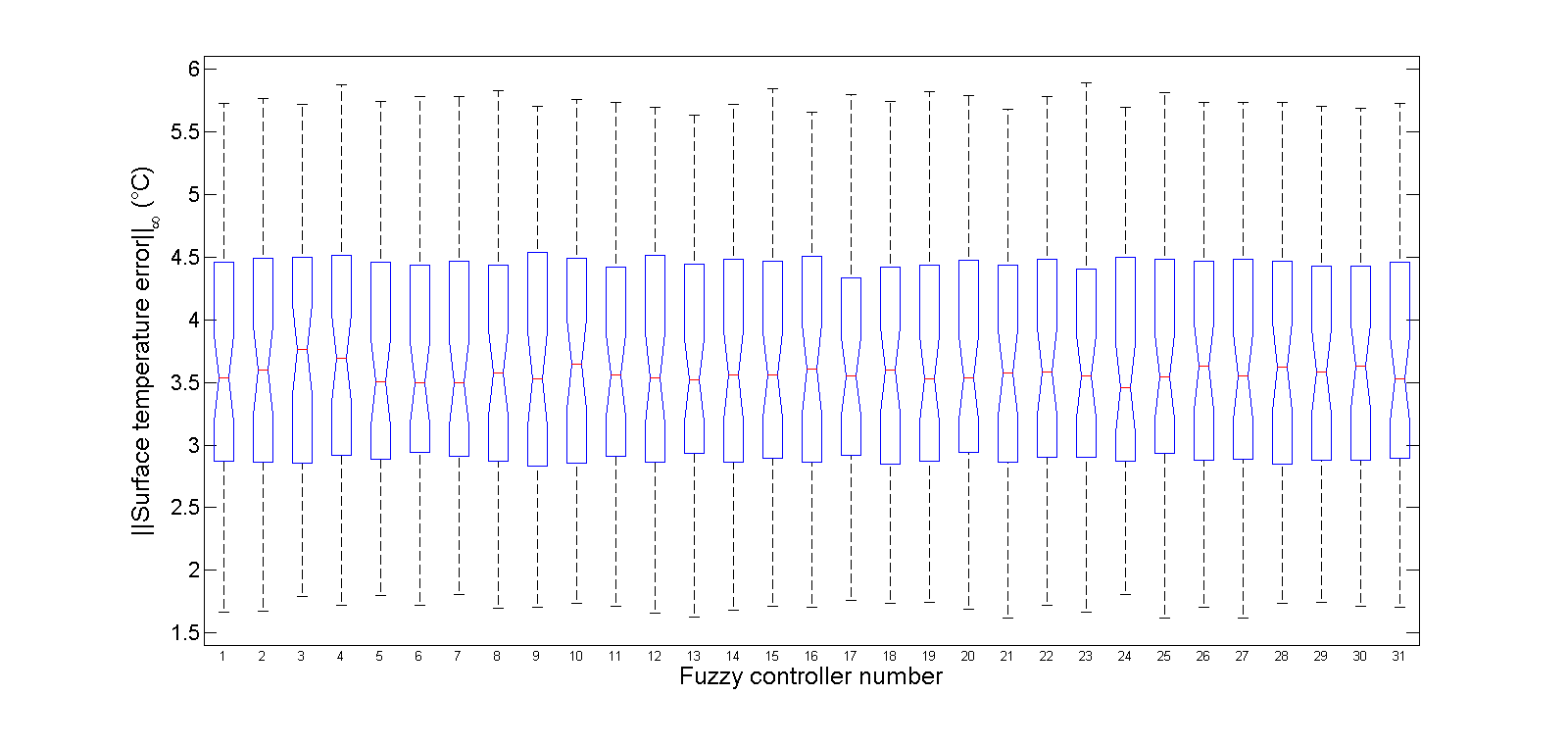}
	\caption{Boxplots of the infinity norm of the temperature error in the case of a disturbed system with noise ($\sigma_{noise}=2^\circ$C) and drift. For the 30 n-f-TILCz (\#1 to \#30) and the i-f-TILC (\#31). For cycles 10 to 60.}
	\label{fig:mousta01}
\end{figure*}

For each situation covered by Tables \ref{ta:no_noise} to \ref{ta:noisy}, a statistical analysis was performed. The Kruskal-Wallis test indicates that all the fuzzy TILCs give similar results, since the hypothesis $\mathcal{H}_0$ is not rejected as the worst $\chi^2$ is 0.4780, which is much lower than the threshold $\chi^2_{0.05;30} = 43.77$.  Figure \ref{fig:mousta01} shows the boxplots of the 30 n-f-TILCs and the i-f-TILC (identified as \#31) corresponding to the situation covered by Table \ref{ta:noisy} with the disturbed model. The similarity of the results is obvious when we look at all the boxplots.

\subsubsection{Case B} 
\begin{figure*}[!t]
	\centering
	\includegraphics[width=\textwidth]{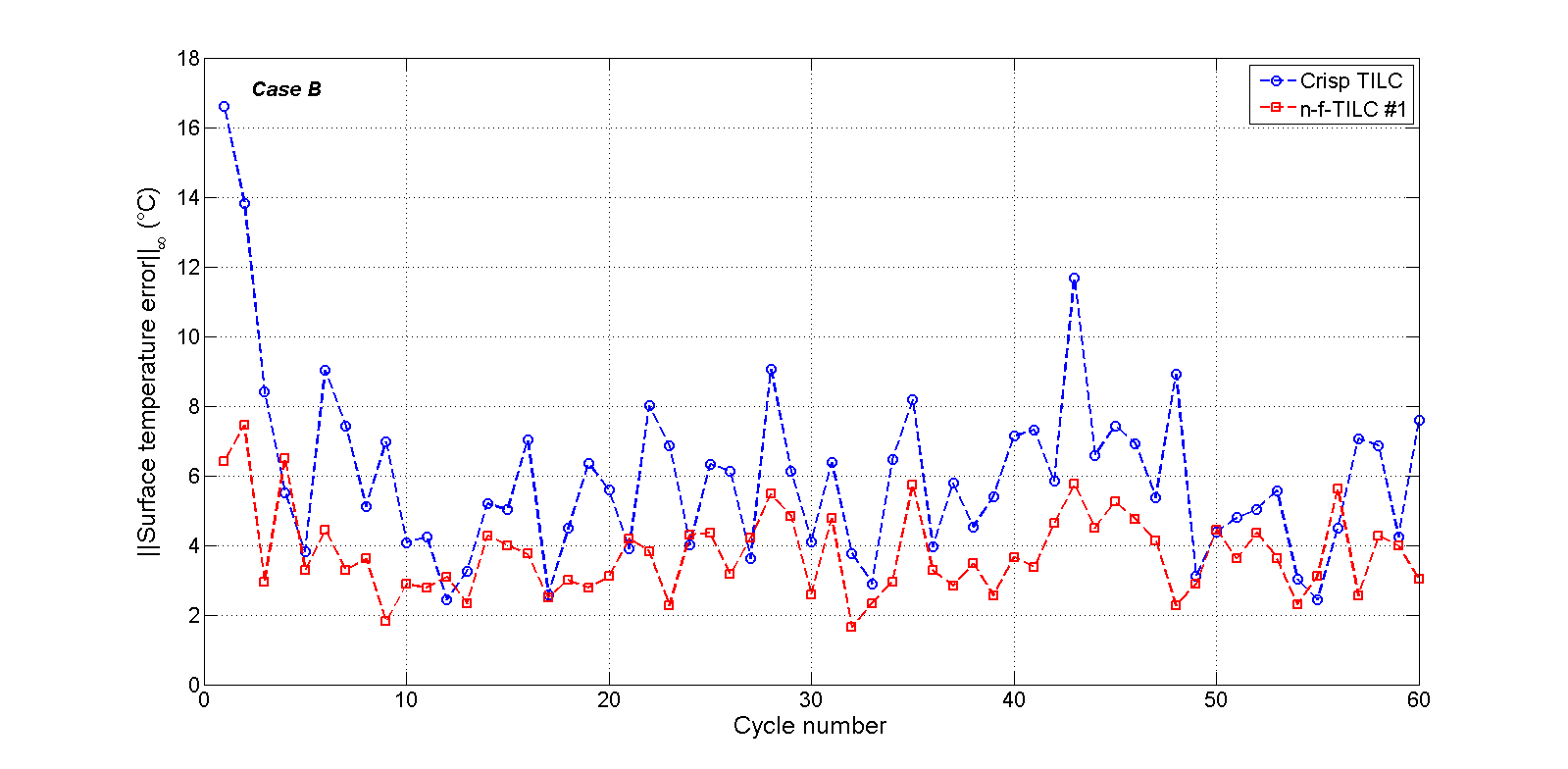}
	\caption{Infinity norm of the surface temperature error in the case of a disturbed system with noise ($\sigma_{noise}=2^\circ$C) and drift.}
	\label{fig:T_error_03}
\end{figure*}

A second set of simulations was performed with a new desired surface temperature vector $\mathbf{y}_d = [140, 140,$ $ 140, 140, 140, 140]^T \: ^\circ$C. When we analyse the results in Table \ref{ta:noisy_caseB} for the situation with noise and drift, we see that the surface temperature errors are in the same order as those obtained in Case A (Table \ref{ta:noisy}). Again, the Kruskal-Wallis test on all the fuzzy TILC results covered by Tables \ref{ta:noisy} and \ref{ta:noisy_caseB} indicate that all the results obtained with the fuzzy TILCs are similar (with $\chi^2 = 1.0472$). Figure \ref{fig:T_error_03} shows the surface temperature error evolution over 60 cycles.

\begin{table}[htbp]
\center
\caption{Simulation results (Case B) --- with noise and drift ($\mu_e$ and $\sigma_e$ for $k \in \{10,\ldots,60\}$)}
\begin{tabular}{|l|l|r|r|r|r|r|}
\hline
TILC algorithm & System & $\mu_e$ ($^\circ$C) & $\sigma_e$ ($^\circ$C) & $||e[1]||_\infty$ ($^\circ$C) \\
\hline
\multirow{2}{*}{Crisp TILC} &
Disturbed & 5.5203 & 1.9141 & 16.6153 \\
& Nominal & 5.3260 & 1.6303 & 14.7302 \\
\hline
\multirow{2}{*}{i-f-TILC} &
Disturbed & 3.6363 & 1.0018 & 10.5150 \\
 & Nominal & 3.6148 & 0.9835 & 6.9217 \\
\hline
n-f-TILC &
Disturbed & 3.6457 & 1.0204 & 10.4140 \\
(mean of 30 controllers) & Nominal & 3.6301 & 0.9949 & 6.8224 \\
\hline
\end{tabular}
\label{ta:noisy_caseB}
\end{table}

The results show that the desired surface temperature seems to have no significant effect on the error level after cycle 10. Again, the heater setpoints for the crisp TILC at cycle $k=1$ are arbitrarily fixed to 350$^\circ$C. Compared to Case A, the initial error is reduced for the crisp TILC, since the desired surface temperature is lower than those requested in Case A.

\begin{table}[htbp]
\center
\caption{Simulation results --- with drift and different levels of noise (All values in $^\circ$C)}
\begin{tabular}{|c|r|r|r|r|r|r|r|r|}
\hline
  & \multicolumn{2}{c}{i-f-TILC} & \multicolumn{2}{|c|}{n-f-TILC ($\sigma_{noise}$)}  & \multicolumn{2}{c|}{"Crisp" TILC} \\
\hline
$\sigma_{noise}$ & $\mu_e$ & $\sigma_e$ & $\mu_e$ & $\sigma_e$ & $\mu_e$ & $\sigma_e$ \\
\hline
1 & 1.8116 & 0.5131 & 1.8117 & 0.5130 & 2.6004 & 0.7985 \\
2 & 3.5901 & 1.0077 & 3.5938 & 1.0122 & 5.2955 & 1.5780 \\
3 & 5.3732 & 1.5737 & 5.3945 & 1.5689 & 8.1212 & 2.4245 \\
4 & 7.1541 & 2.1611 & 8.6887 & 2.8256 & 11.0789 & 3.4452 \\
5 & 8.9642 & 2.7093 & 9.8786 & 3.2403 & 14.2144 & 4.6113 \\
5$^*$ & - & - & 9.5391 & 3.0152 & - & - \\
\hline
\end{tabular}
\label{ta:morenoisy}
\end{table}

\begin{table}[htbp]
\center
\caption{Simulation results --- with different levels of noise (All values in $^\circ$C and no drift)}
\begin{tabular}{|c|r|r|r|r|r|r|r|r|}
\hline
  & \multicolumn{2}{c}{i-f-TILC} & \multicolumn{2}{|c|}{n-f-TILC ($\sigma_{noise}$)}  & \multicolumn{2}{c|}{"Crisp" TILC} \\
\hline
$\sigma_{noise}$ & $\mu_e$ & $\sigma_e$ & $\mu_e$ & $\sigma_e$ & $\mu_e$ & $\sigma_e$ \\
\hline
1 & 1.8405 & 0.5083 & 1.8408 & 0.5085 & 2.6227 & 0.8006 \\
2 & 3.6603 & 1.0521 & 3.6643 & 1.0539 & 5.3443 & 1.5825 \\
3 & 5.4803 & 1.6581 & 5.5022 & 1.6523 & 8.2013 & 2.4334 \\
4 & 7.3448 & 2.2510 & 8.6489 & 2.9075 & 11.1976 & 3.4583 \\
5 & 9.1766 & 2.8496 & 10.0518 & 3.3427 & 14.3811 & 4.6158 \\
5$^*$ & - & - & 9.2790 & 2.9458 & - & - \\
\hline
\end{tabular}
\label{ta:morenoisynodrift}
\end{table}

\subsubsection{Case C} In the last set of simulations, the three TILCs were tested for different noise levels ($\sigma_{noise} \in \{1, 2, 3, 4, 5\}$ $^\circ$C). The results are displayed in Table \ref{ta:morenoisy} when drift is present, and in Table \ref{ta:morenoisynodrift}  Table VIII when there is no drift. In all the simulations, the crisp TILC has a higher temperature error than the two fuzzy TILCs. Apparently, both fuzzy TILCs perform with approximately the same error level.

\begin{figure*}[!t]
	\centering
	\includegraphics[width=\textwidth]{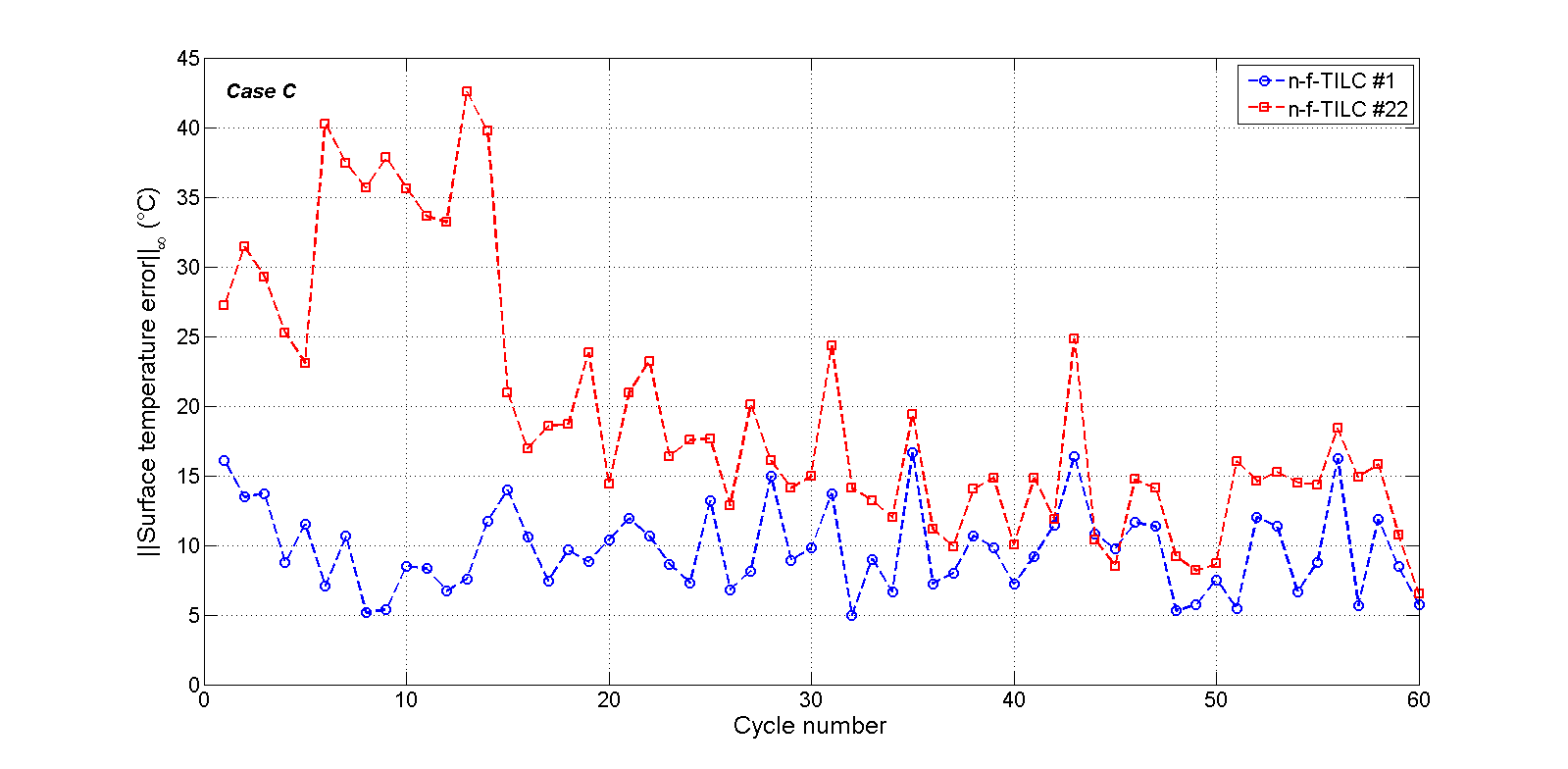}
	\caption{Infinity norm of the surface temperature error in the case of a disturbed system with noise ($\sigma_{noise} = 5$).}
	\label{fig:compare_5}
\end{figure*}

\begin{figure*}[!t]
	\centering
	\includegraphics[width=\textwidth]{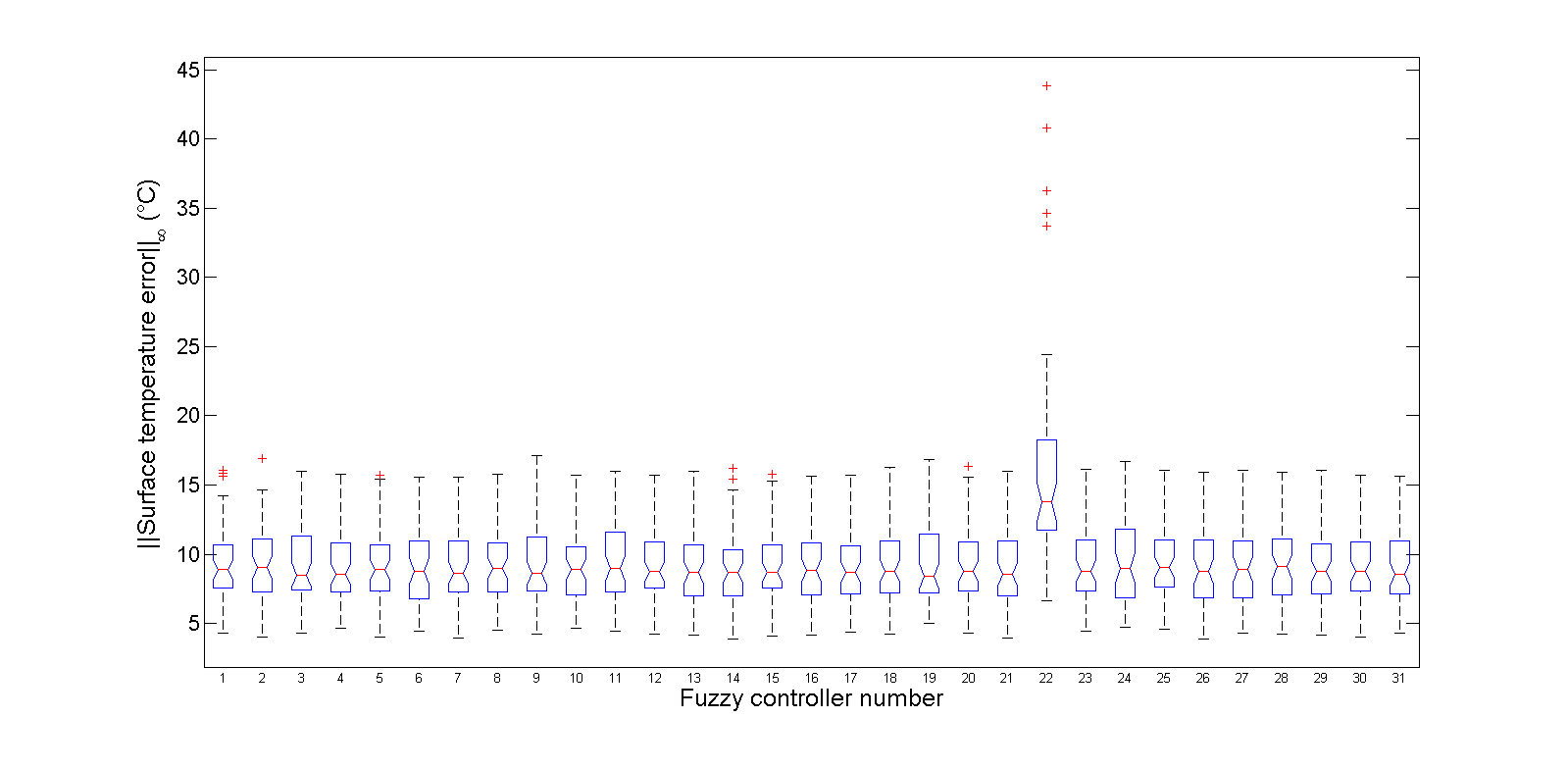}
	\caption{Boxplots of the infinity norm of the temperature error in the case of a disturbed system with noise ($\sigma_{noise}=5^\circ$C) and drift. For the 30 n-f-TILCz (\#1 to \#30) and the i-f-TILC (\#31). For cycles 10 to 60.}
	\label{fig:mousta02}
\end{figure*}

Again using the Kruskal-Wallis test on all the fuzzy TILC results in Tables \ref{ta:morenoisy} and \ref{ta:morenoisynodrift}, we find that we can't reject hypothesis $\mathcal{H}_0$ ($\chi_2$ goes from 0.2614 to 1.3564 and is much lower than $\chi^2_{0.05;30}$), except when $\sigma_{noise} = 5^\circ$C, were hypothesis $\mathcal{H}_0$ is rejected ($\chi_2$ goes from 73.5555 to 90.7538 and is higher than $\chi^2_{0.05;30}$).  For the noisy case (with no drift), there are two outliers in the n-f-TILC, and, for the case with noise and drift, there is one outlier. By removing these outliers, we reduce the error in the case where $\sigma_{noise} = 5^\circ$C, since the remaining results are similar. This corresponds to lines where $\sigma_{noise}$ is identified as $5^*$ in Tables \ref{ta:morenoisy} and \ref{ta:morenoisynodrift}. Figure \ref{fig:compare_5} shows the results for two n-f-TILC algorithms when $\sigma_{noise} = 5^\circ$C. n-f-TILC \#22 is an outlier, while n-f-TILC \#1 is similar to the majority of n-f-TILCs (see the boxplots in Figure \ref{fig:mousta02}).

It is hoped that the noise level ($\sigma_{noise}$) will be somewhere between 1 and 2$^\circ$C on the real process. The simulation results show that, if the measurement signals are very noisy, more than one experiment must be performed for each input tuple, and use the mean value of the output measurements to build the fuzzy model. 

The simulation results also indicate that every measurement signal must be filtered. When the noise level is low, the n-f-TILC seems to be relatively robust to this noise, since the resulting n-f-TILC algorithms behave in a similar way to the i-f-TILC algorithm.

If the 10$^\circ$C target for the error after cycle 10 is easy to obtain with a fuzzy TILC when the noise level $\sigma_{noise}$ is below 3$^\circ$C, the 5$^\circ$C target is obtained only for $\sigma_{noise}$ equal to 1$^\circ$C. This means that it is necessity to filter the measurement signals. The ambient temperature seems to have relatively little effect on the convergence of TILC algorithms in the presence of noise.

\section{Conclusion}\label{se:Conclusion}

This paper has shown the good results obtained using a fuzzy TILC when compared to a 1st order crisp TILC, even in the case where the data used to build the fuzzy TILC were contaminated with a low level of noise.

The proposed design approach provides an easier way to obtain an inverse fuzzy model than the TSK of order 0 for systems with a large number of inputs and outputs, since the fuzzy model relies on linear rules expressed by matrices when all the outputs are combined into a vector. 

The main problem with fuzzy modeling (with a TSK of any order) is the huge number of experiments needed to obtain the data necessary to build the fuzzy model. In a real process, this translates to a significant waste of plastic. With simulation, the only cost is calculation time. 

Since thermoforming machines are used for the production of plastic parts, the data can be obtained from measurements performed on the actual production process. Building a fuzzy model from those measurements is work that we will need to do in the future.

Another potential problem is the effect of noise when the data used to build fuzzy model are obtained from measurements performed on a real process. Optimistically, for the real thermoforming process, the noise level approximately corresponds to the simulations where $\sigma_{noise} = 1^\circ$C. However, noise can become an issue for other processes. Research is needed to analyze how noise can be reduced. A possible solution to this problem is to perform more than one measurement for each input tuple and to use the mean of those measurements as data.

In future work, it will be interesting to analyze how the fuzzy model is obtained and inverted for the case where the number of inputs is not equal to the number of outputs. Moreover, since the simulation results are quite promising, we want to test the fuzzy TILC on an industrial thermoforming machine, as we are convinced that this control can be used safely, without damaging the machine (by overheating).

\section{Appendix \cite{Gauthier2009}}
This appendix is taken from  \cite{Gauthier2009}.

\begin{figure}[htbp]
\centering
\includegraphics[scale=0.3]{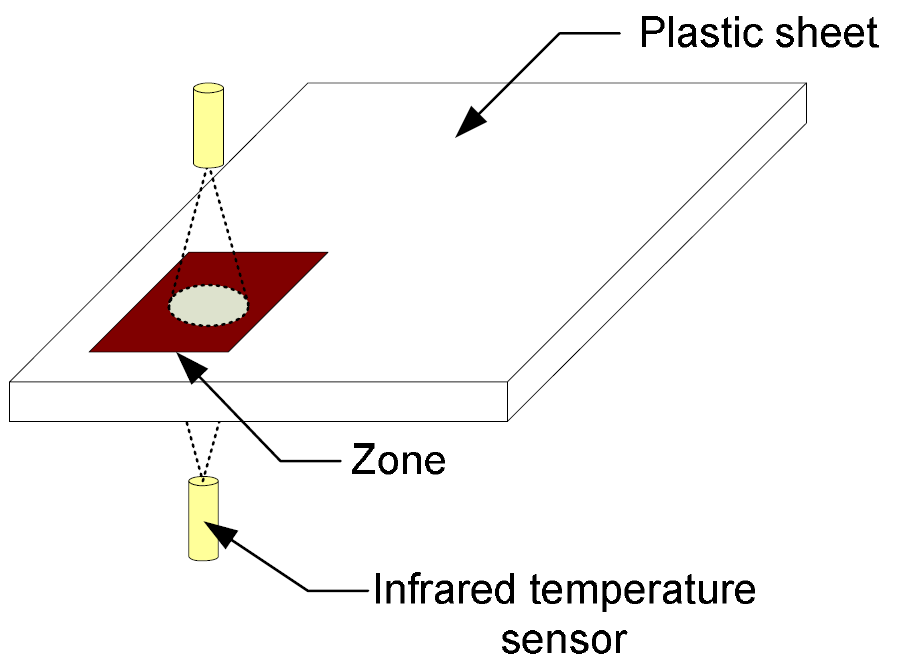}
\caption{IR temperature sensors and the corresponding zone (\cite{Gauthier2008})}
\label{fig:zones}
\end{figure}

The nonlinear model of the thermoforming oven used in the simulations in section VII has been reproduced from the Ph.D. thesis of one of the authors \cite{Gauthier2008}. In the model of the thermoforming oven, the plastic sheet is divided into zones, one for each sensor pair (Figure \ref{fig:zones}).

\begin{figure}[htbp]
\centering
\includegraphics[scale=0.275]{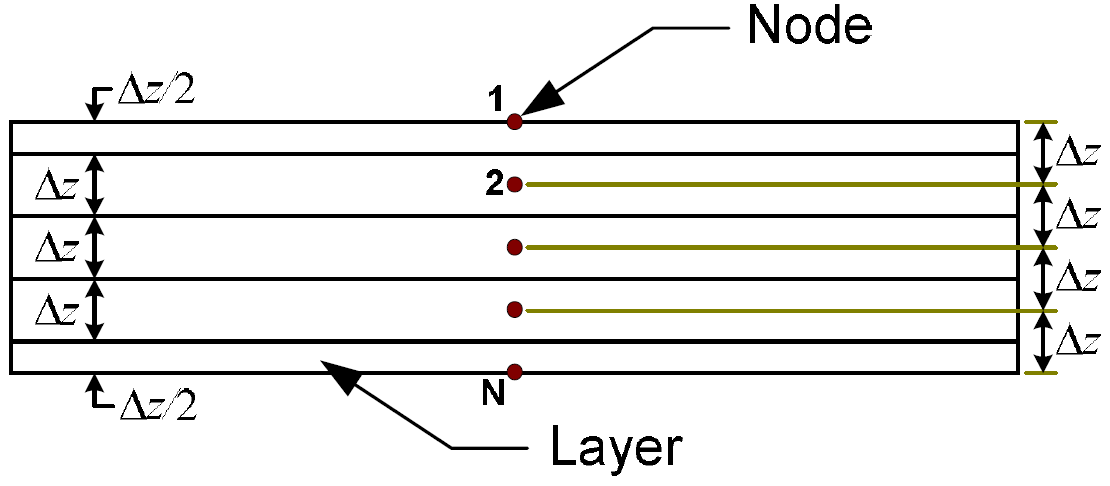}
\caption{Layers and nodes (\cite{Gauthier2008})}
\label{fig:layers}
\end{figure}

To analyze the temperature behavior throughout the plastic sheet, each zone is divided into five layers, and each layer has a node (Figure \ref{fig:layers}). Lateral conduction between nodes of adjacent zones is neglected in this model.

We present the mathematical model first, and then we explain its parameters.

For node 1, in zone $k$ at the top surface of the plastic sheet, the temperature dynamic is expressed by:
\begin{equation}
\frac{d T_{k,1}}{dt} = \frac{2}{\rho V C_p} \left\{ \begin{matrix} \beta_1 Q_{RT_k} + \beta_1 (1-\beta_1)(1-\beta_2)^3 Q_{RB_k} \\ + h \{T_\infty - T_{k,1} \} + \frac{k A}{\Delta z} \{T_{k,2} - T_{k,1} \} \end{matrix} \right \}
\end{equation}

Similarly, for node 5, at the bottom surface of the plastic
sheet:
\begin{equation}
\frac{d T_{k,5}}{dt} = \frac{2}{\rho V C_p} \left\{ \begin{matrix} \beta_1 (1-\beta_1)(1-\beta_2)^3 Q_{RT_k} + \beta_1 Q_{RB_k} \\ + h \{T_\infty - T_{k,5} \} + \frac{k A}{\Delta z} \{T_{k,4} - T_{k,5} \} \end{matrix} \right \}
\end{equation}

For the internal nodes $i$ (of zone $k$), located inside the plastic sheet, we have the following dynamic:
\begin{equation}
\frac{d T_{k,i}}{dt} = \frac{1}{\rho V C_p} \left\{ \begin{matrix} \beta_2 (1-\beta_1) \left\{ (1-\beta_2)^{i-2} Q_{RT_k} + (1-\beta_2)^{4-i} Q_{RB_k} \right \} \\ + \frac{k A}{\Delta z} \{T_{k,i-1} - 2 T_{k,i} + T_{k,i+1} \} \end{matrix} \right \}
\end{equation}

The radiant terms in (43), (44), and (45) are defined for the top heaters as follows:
\begin{equation}
Q_{RT_k} = \sigma \epsilon_{eff} A_h \sum_{j=1}^6 F_{kj} \{ \theta_j^4 - T_{k,1}^4 \}
\end{equation}
and for the bottom heaters as follows:
\begin{equation}
Q_{RB_k} = \sigma \epsilon_{eff} A_h \sum_{j=7}^{12} F_{kj} \{ \theta_j^4 - T_{k,5}^4 \}
\end{equation}

These are the parameters that appear in the Appendix equations:
\begin{itemize}
\item[$\bullet$] $A$ : zone surface, in m$^2$;
\item[$\bullet$] $A_h$ : heater surface, in m$^2$;
\item[$\bullet$] $\beta_1$ : fraction of radiant energy absorbed by a surface
layer;
\item[$\bullet$] $\beta_2$ : fraction of radiant energy absorbed by an
internal layer;
\item[$\bullet$] $C_P$ : specific heat of the plastic sheet, in J/(kg.K);
\item[$\bullet$] $\epsilon_{eff}$ : effective emissivity;
\item[$\bullet$] $F_{kj}$ : view factor between heater $j$ and sheet zone $k$;
\item[$\bullet$] $h$ : convection coefficient, in W/(m$^2$.K);
\item[$\bullet$] k : heat conduction constant, in W/(m.K);
\item[$\bullet$] $\rho$ : density of the plastic sheet, in kg/m$^3$;
\item[$\bullet$] $T_\infty$ : oven ambient temperature, in K;
\item[$\bullet$] $T_{k,i}$ : temperature of node $i$ in zone $k$, in K;
\item[$\bullet$] $\theta_j$ : temperature of heater $j$, in K;
\item[$\bullet$] $\sigma$ : Stefan Boltzmann constant equal to $5.669 \times 10^{-8}$
W/(m$^2$.K$^4$);
\item[$\bullet$] $\Delta z$ : layer thickness, in meters;
\item[$\bullet$] $V$ : layer volume, in m$^3$; ($V = A \Delta z$);

\end{itemize}

Additional details about this model can be found in chapter 2 in \cite{Gauthier2008} or chapter 3 in \cite{Ajersch2004}.


\bibliographystyle{ieeetr}
\bibliography{docu_art}

\end{document}